\documentclass[12pt,preprint]{aastex}

\usepackage{natbib}

\newcommand{\R}{\mathbb{R}}
\newcommand{\bP}{\mathcal{P}}
\newcommand{\bQ}{\mathcal{Q}}
\newcommand{\bK}{\mathcal{K}}
\newcommand{\bS}{\mathcal{S}}
\newcommand{\bD}{\mathcal{D}}
\newcommand{\bA}{\mathcal{A}}
\newcommand{\ith}{$^\mathrm{th}$\ }
\newcommand{\ird}{$^\mathrm{rd}$\ }
\newcommand{\ind}{$^\mathrm{nd}$\ }
\newcommand{\lmin}{{L_\mathrm{min}}}
\newcommand{\lmax}{{L_\mathrm{max}}}
\newcommand{\E}{\mathsf{E}}
\newcommand{\Cov}{\mathsf{Cov}}
\newcommand{\Var}{\mathsf{Var}}
\newcommand{\hmu}{\hat \mu}
\newcommand{\ba}{\begin{eqnarray*}}
\newcommand{\ea}{\end{eqnarray*}}
\newcommand{\mpc}{\frac{\mathrm{km/s}}{\mathrm{Mpc}}}

\slugcomment{Submitted to ApJ, 11/09/06}
\shorttitle{Mapping the Cosmoligical Confidence Ball Surface}
\shortauthors{Bryan et al.}

\begin{document}
\title{Mapping the Cosmological Confidence Ball Surface}
\author{Brent Bryan and Jeff Schneider}
\affil{Department of Machine Learning, Carnegie Mellon University, 5000 Forbes Avenue, Pittsburgh, PA 15213}  
\email{\{bryanba, schneide\}@cs.cmu.edu}

\author{Christopher J. Miller}
\affil{Cerro Tololo Interamerican Observatory, Casilla 603, La Serena, Chile}
\email{cmiller@noao.edu}

\author{Robert C. Nichol}
\affil{Institute of Cosmology and Gravitation, University of Portsmouth, Portsmouth, PO1 2EG, UK} 
\email{bob.nichol@port.ac.uk}

\and
\author{Christopher Genovese and Larry Wasserman}
\affil{Department of Statistics, Carnegie Mellon University, 5000 Forbes Avenue, Pittsburgh, PA 15213} 
\email{\{genovese, larry\}@stat.cmu.edu}

\begin{abstract} 

We present a new technique to compute simultaneously valid confidence
intervals for a set of model parameters. We apply our method to the
Wilkinson Microwave Anisotropy Probe's (WMAP) Cosmic Microwave
Background (CMB) data, exploring a seven dimensional space 
($\tau, \Omega_\mathrm{DE}, \Omega_\mathrm{M}, \omega_{\mathrm{DM}},
\omega_{\mathrm{B}}, f_\nu, n_s$). 
We find two distinct regions-of-interest: the standard
Concordance Model, and a region with large values of $\omega_\mathrm{DM}$,
$\omega_\mathrm{B}$ and $H_0$. This second peak in
parameter space can be rejected by applying a constraint (or a prior)
on the allowable values of the Hubble constant. Our new technique uses
a non-parametric fit to the data, along with a frequentist approach and a
smart search algorithm to map out a statistical confidence
surface. The result is a confidence ``ball'': a set of parameter values that
contains the true value with probability at least $1-\alpha$.
Our algorithm
performs a role similar to the often used Markov Chain Monte Carlo (MCMC),
which samples from the posterior probability function in order to provide
Bayesian credible intervals on the parameters. While the MCMC approach
samples densely around a peak in the posterior, our new technique allows
cosmologists to perform efficient analyses around any regions of interest:
e.g., the peak itself, or, possibly more importantly, the $1-\alpha$ confidence surface.

\end{abstract}

\keywords{cosmology: cosmic microwave background --- cosmology:
 cosmological parameters ---  methods: statistical}

\section{Introduction} \label{sec:introduction}
The Cosmic Microwave Background (CMB) angular temperature power spectrum 
is the most widely utilized data set for constraining the cosmological
parameters \citep{tegmark2001, christensen2001, verde2003, spergel2003, tegmark2004}.
This power spectrum, which
statistically measures the distribution of temperature fluctuations as
a function of scale, is comprised of at least two peaks thought to
have been formed by sound wave modes inherent in the primordial gas during
recombination.  The locations, heights, and height-ratios
of the peaks and valleys in the power spectrum can provide direct
information about fundamental parameters of the universe, such as the space-time
geometry, the  fraction of energy density contained in the baryonic
matter, and the cosmological constant \citep{miller2001}.   However, it
is more common 
for cosmologists to compare the observed CMB power spectrum to a suite  
of cosmological models (e.g. CMBFast \citep{seljak1996} and CAMB
\citep{lewis2000}). These models require as input some minimal number
of cosmological parameters, $d$, --- typically $d=6$ or $d=7$.

Most CMB power spectrum parameter estimations to date have been done via
Bayesian techniques (e.g., \cite{knox2001, gupta2002, spergel2003, jimenez2004, dunkley2005}).
For these techniques, the $d$-dimensional
likelihood function is parametrically estimated and prior probabilities are
assumed for each parameter. Then, a posterior probability distribution
can be computed, and credible intervals can be found.  However, unless
the form of the prior is conjugate on the likelihood (which is atypical), computing the 
posterior involves estimating an integral over the entire space
spanned by the prior.  There are two basic approaches to solving this
problem in the literature.  \cite{tegmark2001} approximates
this integral explicitly, using an adaptive grid, where grid
cells are more densely located in areas presumed to be important.
Secondly, and more popularly, many authors have used Markov Chain Monte Carlo
 (MCMC) (e.g. \cite{gupta2002, lewis2002, jimenez2004, sandvik2004,
dunkley2005,chu2005, hajain2006}), which tend to be much more efficient than grid
based techniques, but are notoriously difficult to tune and test for
convergence \citep{olivestatistics}.   

While Bayesian techniques are used in the majority of work on CMB
parameter estimation, there have also been undertakings to estimate
cosmological parameters using frequentist techniques, such as $\chi^2$ tests
\citep{gorski1993, white1995, padmanabhan2001, 
griffiths2001, abroe2002} and Bayes risk analyses \citep{schafer2003}.
We present a novel frequentist method based upon 
a non-parametric fit to the data to
estimate the smooth underlying power spectrum, as well as an
error ``ellipse'' following the technique used in \cite{miller2001}
and \cite{genovese2004}. 
This confidence ball has a radius which is a function of the
probability with which the true power spectrum is contained within the
ball and the observed error estimates. The ball radius is independent
of both the models to be fit, as well as the parameter ranges to be queried.    Thus, we
can take a vector of parameters, run it through our favorite CMB power
spectrum generating model, and determine whether or not the model (and
hence the parameter vector) lies within our confidence ball, without
fixing \textit{a priori} the model to be used, or the parameter ranges to
be searched.  We are interested in finding the set of parameter
vectors which lie within the $1-\alpha$ confidence  ball, for some
confidence level (or probability of being incorrect), $\alpha$.  

This is a statistically different style of ``confidence'' than the
credible intervals or the ``degree of belief'' one obtains using
Bayesian techniques. In particular, the Bayesian method answers the
question ``assuming a given model and prior distribution over the
parameter space, what is the smallest range of a particular parameter from which I believe
the next sample will be drawn  with probability $1-\alpha$?''  
In contrast, the frequentist approach constructs a procedure for
deriving confidence intervals that when applied to a series of
data sets, traps the true parameters for at least $100(1-\alpha)\%$
of the data sets.   For parametric models with large sample
sizes, Bayesian and frequentist approaches are known to result in
similar inferences.  However, for high dimensional and
non-parametric problems --- such as estimating cosmological parameters
from the CMB power spectrum --- Bayesian methods may not yield accurate  
inferences \citep{olivestatistics}.  In such cases, the Bayesian 95\%
credible interval may not contain the true value 95\% of the time in
a frequency sense.   

Additionally, mapping a region of high likelihood points in
parameter space is fundamentally a search problem.  As MCMC methods are
designed to sample and/or integrate a distribution, they are not
necessarily good search algorithms in practice.  In particular, a MCMC
method ``represents'' a high-likelihood region by heavily sampling
that region --- an expensive proposition when using CMBFast.  In
contrast, a search algorithm that can directly observe the 
(normalized) likelihood of a sample will have no reason to spend more
samples in the same location.
In addition to describing a frequentist approach to computing
confidence intervals for cosmological parameters, 
another significant contribution of this paper is the
proposal of a new search algorithm for mapping confidence surfaces.

\begin{figure*}
\begin{center}
\plotone{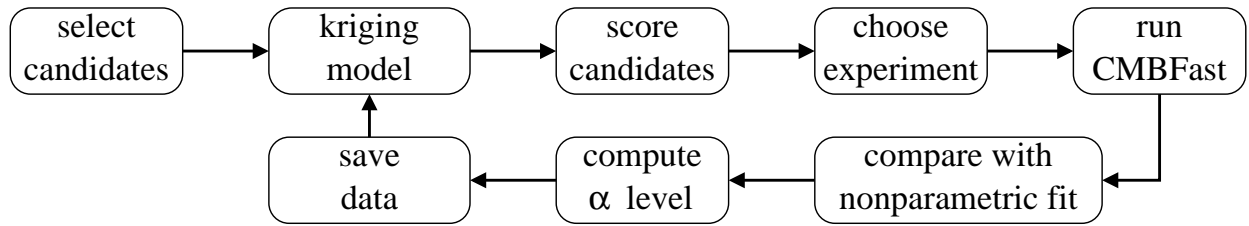}
\end{center}
\caption{Schematic outline of our technique to constraint confidence intervals.}
\label{fig:outline}
\end{figure*}

In this work, we utilize the non-parametric basis
described by  \cite{miller2001} and \cite{genovese2004} to constrain
the set of cosmological models which fit the WMAP observations. 
At the same time, we must deal with the challenges posed in other
frameworks namely: robustness of the algorithm, efficiency,
and issues of convergence.  A schematic outline of our technique is
shown in Figure \ref{fig:outline}.
In \S \ref{sec:methodology}, we
briefly describe the data and cosmological models used, as well as the
non-parametric technique (the bottom row of Figure \ref{fig:outline}).  We 
then focus on a new algorithm to map the derived confidence ball into
parameter space in \S \ref{sec:algorithm}, sketched out on the
top line of Figure \ref{fig:outline}.   In \S
\ref{sec:results}, we present results of our
algorithm, and discuss challenges to accurately determine confidence
intervals using any statistical approach.
Finally, in \S \ref{sec:comparison}, we compare our
method with commonly used inference techniques, and discuss the
advantages of using the proposed approach.

\section{Methodology} \label{sec:methodology}
\subsection{Data \& Models} \label{sec:datamodels}
We examine the CMB power-spectrum ($\hat C_{\ell}$) as
measured by the Wilkinson Microwave Anisotropy Probe's first-year
data release \citep{bennett2003, hinshaw2003,
verde2003}\footnote{Available at 
\url{http://lambda.gsfc.nasa.gov}}, shown in Figure
\ref{fig:wmapdata1}.  Our approach is similar to that of other authors
(e.g. \cite{tegmark1999, tegmark2001,  
spergel2003}), who fit the observed CMB power spectrum to a suite of
cosmological models.   These models, while sophisticated and detailed, have numerous free
parameters, some of which are difficult to ascertain (e.g. ionization
depth, contribution of gravity waves).  However, there are many codes
available to compute CMB power spectrum, which trade off speed for
accuracy and robustness. 

Both CMBFast \citep{seljak1996} and the related CAMB \citep{lewis2000}
compute the CMB power spectrum by evolving the Boltzmann equation
using a line of sight integration technique.  While an order of
magnitude faster than computing the full Boltzmann solution, this approach
is still rather slow.
One approach for reducing the computation time of CMBFast
is to split the Boltzmann computation into low and
high multipole moment portions, as the low and high multipoles are
mostly independent \citep{tegmark2001}.  Using this method, ksplit,
\cite{tegmark2001} was able to reduce computation time by a factor of 10.
Additionally, several approximate programs have been
developed which are orders of magnitudes faster than CMBFast,
including DASh \citep{kaplinghat2002},
CMBWarp \citep{jimenez2004}, and Pico \citep{fendt2006}.  
In general, these programs gain great speedups 
by approximating the power spectrum with a regression function
fit to predetermined sample points generated from simulators such as CMBFast.  As a
result, generating a hypothesis spectrum for a new set of parameters
is a simple function evaluation, foregoing the computation of the
Boltzmann equation entirely.  

While using any one of these approximate methods or ksplit may seem
appealing due to their computational efficiency, they do not have the
desired accuracy and robustness \citep{seljak2003}.  These codes are
only approximations.  While fairly accurate around the concordance
peak, their accuracy drops off drastically when computing models for
parameter vectors slightly removed from
the ``accepted'' cosmological models.  
Additionally, these codes are prone to failures when presented with
parameter vectors that are not within a narrowly defined region around
the concordance model \citep{fendt2006}.
According to the Pico website: ``Since Pico's purpose is to be part of
parameter estimation codes, we are mainly concerned with having the regression
coefficients defined around the region of parameter space allowed by
the data (mainly the WMAP3 data). Pico will not be able to compute
accurate spectra and likelihoods away from this region, but it will
warn you about this.''  Similarly, in many instances ksplit will hang
on parameter vectors that are a short distance from the concordance peak.
Since we are interested in finding the tightest possible confidence
intervals for all regions of parameter space that can possibly fit the data, 
we do not want to be artificially restricted by our CMB simulator.
Thus, we choose to compute the model CMB power spectra
using CMBFast; while not the fastest code available CMBFast is
accurate and reliable. 

Next, multipole covariance is estimated by using the covariance derived for the
concordance model using code from \cite{verde2003}.  
We find that the computed variances match well with
those found in the first-year data release, with only a slight
(roughly $1.15$) multiplicative offset.  This constant factor offset
was hinted at by the sub unity slope of the quantile-quantile plot of
the variance weighted deviations between the data and the concordance
model prediction, using the variances given in the WMAP data.

\begin{table}[t]
\begin{center}
\begin{tabular}{c  l r@{${\ }-{\ }$}l}
\hline
\textbf{Parameter} & \textbf{Description} &
\multicolumn{2}{c}{\textbf{Range}}\\
\hline
\hline
$\tau$ & optical depth & 0.0 & 1.2\\
$\Omega_\mathrm{DE}$ & dark energy mass fraction& 0.0 & 1.0\\
$\Omega_\mathrm{M}$ & total mass fraction & 0.1 & 1.0\\
$\omega_{\mathrm{DM}}$ & dark matter density& 0.01 & 1.2 \\
$\omega_{\mathrm{B}}$ & baryon density& 0.001 & 0.25\\
$f_\nu$ & neutrino fraction& 0.0 & 1.0\\
$n_s$ & spectral index& 0.5 & 1.7\\
\hline
\end{tabular}
\end{center}
\caption{Cosmological parameters and ranges searched.}
\label{paramtable}
\end{table}

\cite{spergel2006} show that the WMAP third year data are well
described by a simple 6 parameter model: $\tau, H_0,\Omega_\mathrm{M},
\Omega_\mathrm{B}, \sigma_8, n_s$.  In this paper, we use
effectively the same model space as the simplified model in
\cite{spergel2006}, except that we include the neutrino fraction and
exclude $\sigma_8$.  We made this change as we are not utilizing
large-scale structure data, which is sensitive to $\sigma_8$. The
resulting parameter vector 
$\mathbf{p} = (\tau, \Omega_\mathrm{DE}, \Omega_\mathrm{M},
\omega_{\mathrm{DM}}, \omega_{\mathrm{B}}, f_\nu, n_s)$ is similar to
the model space searched by \cite{tegmark2001}.
 A description and considered range for each of these variables
is presented in Table \ref{paramtable}; the parameter ranges
considered here are slightly larger than those searched by \cite{tegmark2001}, due
to our interest in mapping an observed secondary peak in parameter space.  
Note that $\Omega_\mathrm{k} = 1 - \Omega_\mathrm{M} - \Omega_\mathrm{DE}$.
Moreover, the Hubble constant, $H_0$, is not an independent parameter,
but given by
\[
\frac{H_0}{100} = h =\sqrt{\frac{\omega_\mathrm{DM}+\omega_\mathrm{B}}{\Omega_\mathrm{M}}}
= \sqrt{\frac{\omega_\mathrm{DM}+\omega_\mathrm{B}}{1-\Omega_\mathrm{k} - \Omega_\mathrm{DE}}}.
\]
We denote the space spanned by $\mathbf{p}$ as
$\mathcal{P}$. $\mathcal{P}$ is a seven dimensional hyper-rectangle
where the range of the $j^\mathrm{th}$ side corresponds to the range of the
$j^\mathrm{th}$ cosmological parameter of $\mathbf{p}$.  

\subsection{Nonparametric Analysis} \label{sec:nonparametric}
We now provide a brief sketch of nonparametric data
analysis, as it pertains to the CMB power spectrum.  We follow the
derivations given in \cite{miller2001} and \cite{genovese2004}, and refer
interested readers to those works. Our technique is
designed to:
\begin{enumerate}
\item Compute a fit to the actual data which minimizes the sum of the
      bias and the variance between the fit
      and the data, taking into account the full covariance discussed in
      \S \ref{sec:datamodels}.  Errors are assumed to be Gaussian.
      This fit is effectively a smoothed version of the data.  

\item Determine a confidence ellipse ball around the best fit for a
      given test level, $\alpha$.

\item Find all such vectors $s \in \mathcal{P}$ such that the
      power spectrum output by CMBFast for $s$ results in a 
      model which is contained within the $1-\alpha$ confidence ball
      found in step 2.  
\end{enumerate}
We now detail items 1 and 2, leaving the discussion of item 3 to \S \ref{sec:mapping}.

%
%
%

\subsubsection{The Non-Parametric Fit} \label{sec:fit}
Let $\ell \in [L_\mathrm{min}, \dots, L_\mathrm{max}]$ denote a
generic index of the CMB temperature power spectrum multipole, and $n
= L_\mathrm{max} - L_\mathrm{min}+1$ be the total number of observed
multipoles. We take $Y_{\ell} = \hat{C}_{\ell}$ to be the observations
of the CMB where  $x_{\ell} = (\ell-\lmin)/(\lmax-\lmin)$ and let 
$f(x_{\ell})\equiv C_{\ell}$ denote the true power spectrum
at multipole index $\ell$. 
We then solve the nonparametric regression problem:
\begin{equation}\label{eq:regress2}
Y_{\ell} = f(x_{\ell}) + \epsilon_{\ell}, \qquad \ell = L_\mathrm{min}, \ldots, L_\mathrm{max},
\end{equation}
where $\epsilon =
(\epsilon_{L_\mathrm{min}},\ldots,\epsilon_{L_\mathrm{max}})$ are
assumed Gaussian with known covariance matrix $\Sigma$ as described earlier.
Henceforth, we will use $i=\ell -\lmin+1$ as an index.
Nonparametric analysis is based on the notion of estimating a function
without forcing it to fit some finite-dimensional parameter form
(e.g. a Normal distribution), by smoothing the data in such a way to
balance the bias and variance. In this work, we use orthogonal series
regression to estimate $f$, expanding $f$ as a
cosine basis: 
\[
f(x) = \sum\limits_{j=0}^\infty \mu_j \phi_j(x)
\]
where
\[
\phi_j(x) =
\left\{
\begin{array}{l l}
1 & \mathrm{for\ } j=0\\
\sqrt{2}\cos(\pi j x) & \mathrm{for\ } j = 1,2,3, \dots
\end{array}
\right.
\]
and the $\mu_j$'s are the coefficients for each basis component.
If $f$ is smooth, then $\mu_j$ will decay rapidly as $j$
increases. That is, if $f$ is smooth, then there are little or no
high frequency fluctuations in $f$ and hence $\mu_j \simeq 0$.
Thus,
$\sum_{j=n+1}^\infty \mu^2_j$ will be negligible, and we can approximate the
infinite sum as $f(x) \approx \sum_{j=0}^n \mu_j \phi_j(x)$.  Let 
\[
Z_j = \frac{1}{n} \sum_{i=1}^n Y_i \phi_j(X_i)
\]
for $j=0, 1,\dots n$.  Then
$Z$ is approximately normal distributed with mean $\mu$ and covariance
$B/\sqrt{n} = U \Sigma U^T/ \sqrt{n}$, where $U$ is the cosine basis transformation
matrix.  

 In order to obtain an even smoother
estimate of $f$, we damp out the higher frequencies using shrinkage
estimators.  We let $\hat \mu_j = \lambda_j Z_j$ where $1 \ge
\lambda_0 \ge \lambda_1 \ge \cdots \ge \lambda_n \ge 0$ are shrinkage
coefficients.  The estimate of $f$ is now
\[
\hat f(x) = \sum_{j=0}^n \hat \mu_j \phi_j(x) = \sum_{j=0}^n 
\lambda_j Z_j \phi_j(x).
\]
Following \cite{genovese2004}, we use a special case of monotone
shrinkage in which 
\[
\lambda_j = \left\{ 
\begin{array}{cc}
1 & \mathrm{for\ } j\le J\\
0 & \mathrm{for\ } j> J
\end{array}\right.
\]
for some integer $J \in [0,n]$.  We will show how to find $J$ shortly.
Using the monotone shrinkage scheme described above, the estimate of $f$ becomes
\[
\hat f(x) =  \sum_{j=0}^J  Z_j \phi_j(x).
\]

The squared error loss as a function of $\hat \lambda = (\hat \lambda_0,\hat
\lambda_1,  \dots, \hat \lambda_n)$ is
\[
L_n(\hat \lambda) = 
\int_0^1 \left(\frac{\hat f(x) -
  f(x)}{\sigma(x)}\right)^2 \, dx\approx \sum_{j=1}^n
\left(\frac{\mu_j - \hat \mu_j}{\sigma_j}\right)^2,
\]
where $\sigma^2(x)$ is the variance of $f$, and $\sigma_j^2$ are the
observed variances of the power spectrum (the elements on the diagonal
of $\Sigma$).  Meanwhile, the risk is given by
\[
R(\lambda) = \E \left[\int_0^1 \left(\frac{\hat f(x) -
  f(x)}{\sigma(x)}\right)^2 \, dx \right] \approx 
\frac{J}{n} + \sum_{j=J}^n \frac{\mu_j^2}{\sigma_j^2}
\]

We choose $J$ to minimize the Stein's unbiased risk estimate
\begin{equation}\label{eqn:stein}
\hat R = Z^T \bar D W \bar D Z + \mathrm{trace}(DWDB) -
\mathrm{trace}(\bar D W \bar D B)
\end{equation}
where $D$ and $\bar D = 1 -D$ are diagonal matrices with 1's in the
first $J$ and last $n-J$ entries respectively, $B$ is the covariance
of $Z$, and $W_{jk} = \sum_{\ell} \Delta_{jk\ell}/\sigma_\ell$ and 
\begin{eqnarray*}
\Delta_{jk\ell} &=& \int_0^1 \phi_j \phi_k \phi_\ell\\
&=& \left\{ 
\begin{array}{c c}
1 & \mathrm{if\ \#}\{j,k,l = 0\} = 3\\
0 & \mathrm{if\ \#}\{j,k,l = 0\} = 2\\
\delta_{jk}\delta_{0\ell} + \delta_{j\ell}\delta_{0k} +
\delta_{k\ell}\delta_{0j} & \mathrm{if\ \#}\{j,k,l = 0\} = 1\\
\frac{1}{\sqrt{2}}(\delta_{\ell, j+k} + \delta_{\ell,|j-k|}) & \mathrm{if\ \#}\{j,k,l = 0\} = 0
\end{array}
\right..
\end{eqnarray*}
\cite{beran1998} showed that $\hat R(\lambda)$ is asymptotically,
uniformly close to $R(\lambda)$ when using monotone shrinkage
coefficients and $\sigma(x)=1$.  \cite{genovese2004} extended this
result to the heteroskedastic case used here.
 
\begin{figure}
\begin{center}
\noindent
\plotone{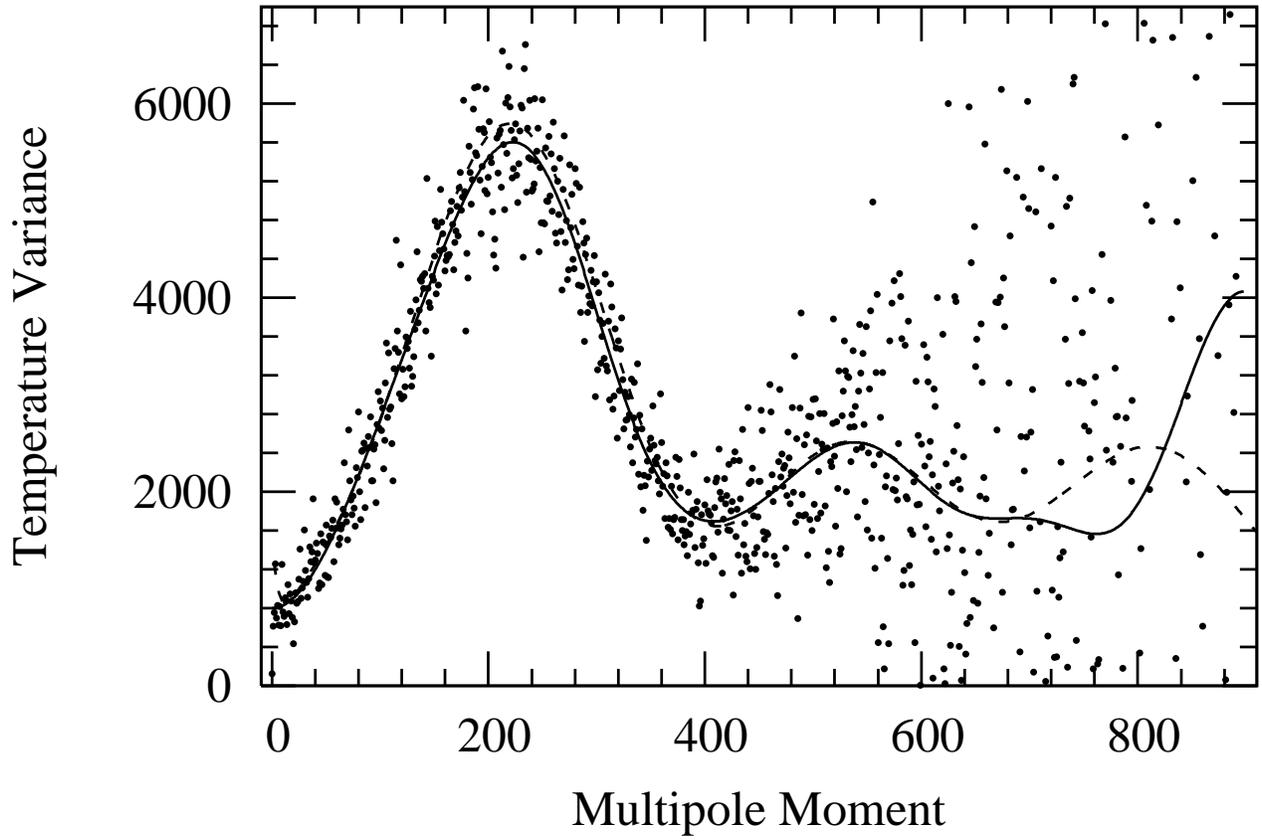}
\end{center}
\caption{Comparison of our nonparametric fit of the CMB power-spectrum
(solid) with \cite{spergel2003} parametric fit (dashed).  First-year
 WMAP data (dots) are shown without errors for clarity.}
\label{fig:wmapdata1}
\end{figure}

In Figure \ref{fig:wmapdata1}, we compare our non-parametric 
fit to the WMAP data to a model-based fit from \cite{spergel2003}.
Points in the figure depict the first year WMAP data. 
Error bars are omitted for clarity.  The full estimated
covariance, $\Sigma$, is used in both the \cite{spergel2003} model fit
and the \cite{genovese2004} non-parametric fit.  

\subsubsection{The Confidence Ball} \label{sec:confball}
After we perform the non-parametric fit,
we need to quantify the uncertainty to make statistical inferences.
We use the Beran-D\"umbgen pivot method \citep{beran1998,beran2000} to
derive valid confidence intervals.  This method relies
on the weak convergence of the ``pivot process'' --- $B_n(\hat \lambda) =
\sqrt{n} (L_n(\hat \lambda) - \hat R (\hat \lambda))$ --- to a Normal
$(0, \tau^2)$ distribution for some $\tau^2 >0$; a derivation of $\hat
\tau_n$ can be found in Appendix \ref{appendix}, taken from Appendix 3 of 
\cite{genovese2004}.  Using the convergence of the pivot process, we can 
compute a confidence ellipse for the basis coefficients with a
``radius'' given by:
\begin{equation} \label{conf0}
\mathcal{D}_n
  = \left\{\mu : \sum_{i=1}^n \left(\frac{\hat{\mu}_i - \mu_i}{\sigma_i}\right)^2 \le
 \frac{\hat\tau_n \, z_\alpha}{\sqrt{n}} + \hat{R}(\hat\lambda_n)\right\}
\end{equation}
where the best fit to the data is represented by
$\hat{\mu}_i$, the function being tested (whether it is within some
confidence ball) is $\mu_i$, and the level of the confidence ball is
determined by $z_\alpha$, the upper $\alpha$ quantile of a standard
Normal distribution.  

Therefore, using the central limit theorem, we have
\begin{equation} \label{conf}
\mathcal{B}_n = \left\{f(x) = \sum_{j=0}^n \mu_j \phi_j(x): \mu \in
\mathcal{D}_n \right\} 
\end{equation}
is an asymptotic $1-\alpha$ confidence set for $f$.  

Thus, to determine if any
given vector $s \in \mathcal{P}$ is within our confidence ball, we
merely have to run our cosmological model to compute the resulting
power spectrum, 
$\hat f(s)$, and check to see if $\hat f(s) \in \mathcal {B}_n$.  This can 
be easily done by using Equation \ref{conf0} to check whether the sum of 
squares of $\hat \mu$ and $\mu$ are less than a constant given on the
right-hand side of Equation \ref{conf0}.
As shown in Figure \ref{fig:distance_alpha}, as the radius increases,
so does the size of the confidence set (and $\alpha$ decreases). Thus,
a 95\% (or $\alpha = 0.05$) confidence region has a larger ``radius''
than does a 67\% (or $\alpha = 0.33$) confidence region.  Moreover, a
$1-\alpha$ confidence ball strictly contains
all confidence balls with smaller values of $1-\alpha$.

\begin{figure}
\begin{center}
\plotone{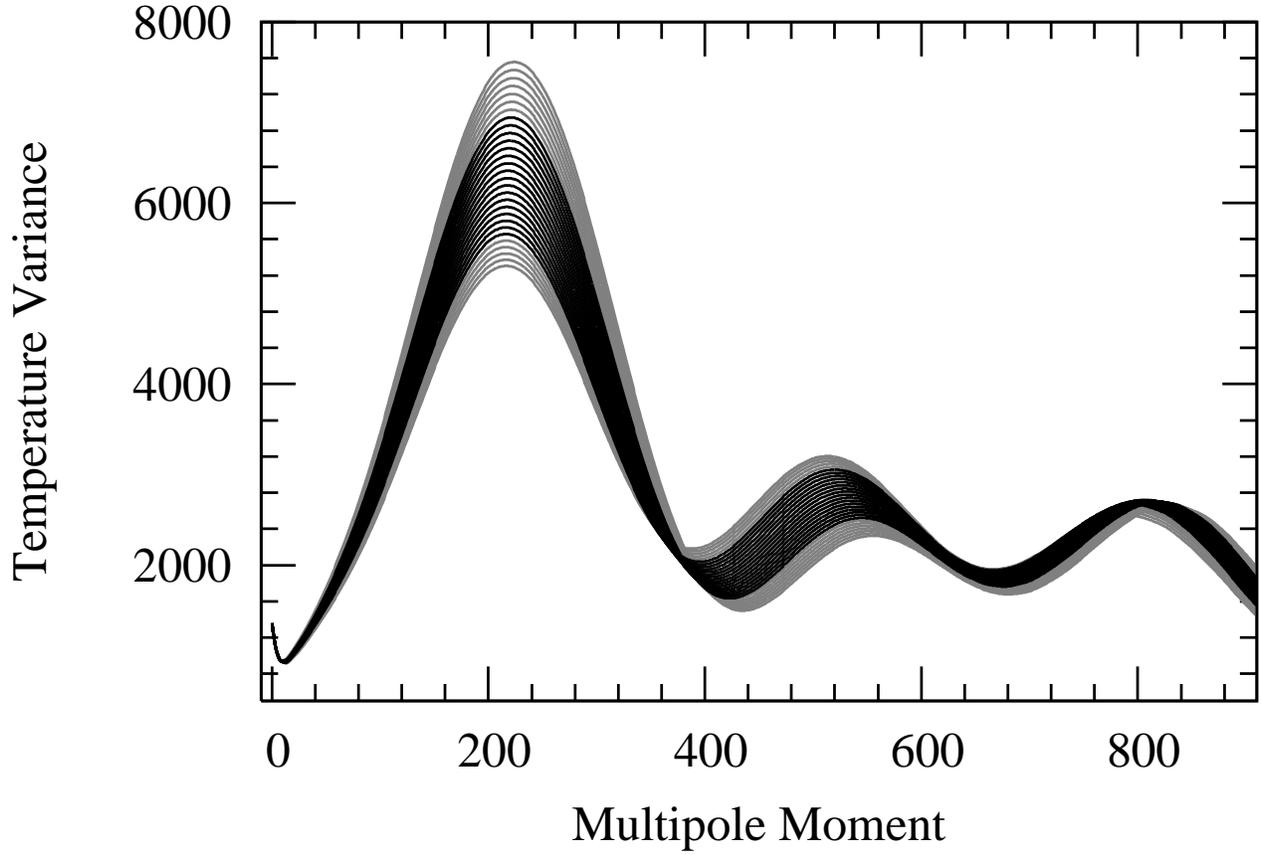}
\end{center}
\caption{A ``ribbon'' plot depicting the effect of varying
  $\omega_\mathrm{B}$ while all other parameters remain fixed (at
  concordance values).  Black lines indicate those models which are
  contained within a 95\% confidence ball, while gray lies indicate
  those models rejected by the hypothesis that the model and the
  regressed fit are the same.}
\label{fig:wmapdata2}
\end{figure}

Since the dimensionality of our space is large, it is difficult to
visualize the confidence region that surrounds the non-parametric fit.
However, we can show examples of functions which live inside (or outside)
our confidence region by calculating their distance from the
nonparametric fit to the data.
In Figure \ref{fig:wmapdata2}, we show a ``ribbon'' plot for
$\omega_\mathrm{B}$ around the concordance model.  This figure is generated by
setting all of the cosmological parameters to their concordance
values and then slowly evolving $\omega_\mathrm{B}$ from $0.012250$ to
$0.036750$ to depict the range of temperature spectra allowed due to
uncertainty of $\omega_\mathrm{B}$.  The
black curves are cosmological models which live within the
$95\%$ confidence ball, while gray curves are models that do not.
As can be seen in this figure, the shape of the confidence region is
not simply a band of constant width surrounding the best fit. It is, in fact,
a very complicated, possibly disconnected surface in our high-dimensional
parameter space. {\it It is this confidence surface that we wish to map in detail.}

\section{Mapping the Confidence Surfaces} \label{sec:mapping}
While theoretically Equation \ref{conf} exactly gives us the $1-\alpha$
confidence bound for any functional of the data, it is not trivial to
compute what these bounds are.  While it is easy to use Equation
\ref{conf} to compute whether or not a given model is within the
confidence ball, the method outlined in \S \ref{sec:nonparametric}
does not provide a way to easily compute all those spectrum that lie
within that ball.

Concretely, when we test if a CMB power
spectrum lies within the ball, we compare the given spectrum with
the non-parametric fit found above, by computing a variance weighted
sum of squares between the given spectrum and the regressed model.  We
call this weighted sum of squares the test spectrum's ``distance''.  If
we are given a model which results in a test spectrum whose distance
is greater than the radius of our confidence ball, then we can reject
the test spectrum (and its associated parameter vector) at the
$1-\alpha$ level. 
If not, then our test does not have the power to distinguish between
the regressed model and our test model.  Note that we are taking a
$\sim900$ element spectrum and compressing it to a scalar.  Thus, there
are many models --- possibly representing vastly different spectra ---
that may result in exactly the same distance value. For the hypothesis
test that the fitted function and regressed models are derived form the same 
distribution, we will draw the same conclusion for all models with the
same distance values.  Either all models with a particular distance
score can be rejected or none can.  For a
given confidence ball radius, we could compute (possibly with some discrete
approximation) all of the possible CMB power spectra that have
distances equal to the confidence radius.  However, we are unaware of
an easy way to determine the cosmological parameters of a power
spectrum given only the power spectrum itself.  That is, we do not have a
method to easily invert CMBFast.

%

Of course, one solution would be to grid the parameter space, and
run a model for each grid cell.  We could then use these models to
approximate the mapping between parameter vectors and confidence
level using, for instance, a simple linear approximator.
As noted in \S \ref{sec:introduction}, such an approach
is far too slow, explaining why \cite{tegmark2001} use both
adaptive grids and a modified version of CMBFast.   
Instead, we suggest an adaptive approach, which allows us to determine
confidence intervals of our cosmological parameters more quickly and
accurately.   
In particular, we are able to quickly refine our approximating surface 
in the areas of interest -- those near the confidence ball's radius --
while ignoring the uninteresting regions.  This allows us to obtain
estimates of the $1-\alpha$ confidence intervals of our
cosmological parameters much more efficiently.

\subsection{Modeling Known Experiments} \label{model}
The combination of CMBFast and the confidence ball method gives us a scoring
function  $f:\mathcal{P}\to \R$, which takes an input vector of
parameters ($s \in \mathcal{P}$) and returns a distance value. This is
accomplished by plugging the cosmological parameter values of $s$ into
CMBFast to compute a model power spectrum, and then comparing this
model spectrum with our non-parametric fit to the observed power spectrum
using Equations \ref{conf0} and \ref{conf}.
Given a particular $1-\alpha$ confidence ball radius, $t$, 
we want to find the set of points, $\bS$ ($\bS \subseteq \bP$), that have
distances to the regressed fit of the data less than or equal to the
confidence ball radius: $\{s \in \bS | s \in \bP, f(s) \le t\}$.   
Since we can not easily invert $f$ --- that is to say CMBFast ---
we must deduce $\bS$ by carefully sampling the points in $\bP$.

For CMBFast, the cost to compute $f(s)$ given $s$ can be significant:
computing power spectra away from the concordance model can take 
5 to 15 minutes.
Thus, care should be taken when choosing the
next experiment, as picking optimum points can reduce the run time of
the algorithm by orders of magnitude.  Thus, it is preferable to
analyze current knowledge about the underlying function and select experiments
which quickly refine the estimate of the distance function around the
confidence ball radius.  There are several methods one could use to
create a model of the data, notably some form of parametric
regression.  However, we chose to approximate $f(s)$ using
Gaussian process regression, as other forms of regression may
smooth the data, ignoring subtle features of the function that may
become pronounced with more data.  A Gaussian process is
a non-parametric form of regression.  Predictions for
unobserved points are computed by using a weighted combination of the
function values for those points which have already been observed,
where a distance-based kernel function is used to determine the
relative weights.  These distance-based kernels generally weight nearby points
significantly more than distance points.
Thus, assuming the underlying function is continuous,
Gaussian processes will perfectly describe the function given an
infinite set of unique data points.

In this work, we use ordinary kriging, a form of Gaussian processes that 
assumes that the semi-variance, $\mathcal{K}(\cdot, \cdot)$, between
two points is a linear function of their distance \citep{cressie1991};
for any two points $s_i, s_j \in \bP$, 
\[
\mathcal{K}(s_i, s_j) = \frac{k}{2} \E\left[ \Big(f(s_i) - f(s_j)\Big)^2\right]
\]
where $k$ is a constant --- known as the kriging
parameter --- which is an estimate of the maximum magnitude of the
first derivative of the function.  Therefore, the
expected semi-variance between two points, $s_i, s_j \in \bP$ is given
by
\ba
\gamma(s_i, s_j) &=& E(\bK(s_i, s_j)) = k \bD(s_i, s_j)+c\\
&=&k \left[\sum\limits_{\ell=1}^d \alpha_\ell^2(s_{i\ell} - s_{j\ell})^2\right]^{1/2}+c
\ea
where $\bD(\cdot, \cdot)$ is a distance function defined on the parameter
space $\bP$ and $c$ is the observed variance (e.g. experimental noise) 
when repeatedly sampling the function $f$ at the same location.  
We have found that using a simple weighted 
distance function where each dimension is linearly scaled by the
parameter $\alpha_\ell$, as depicted in the previous equation,
reasonably ensures that parameters are given equal 
consideration given their disparate values and derivatives.  For our
analysis, we adjusted the $\alpha_\ell$'s to ensure that the maximum derivative
along each dimension was approximately 1 during the sampling process.
Additionally, while the simulations computed by CMBFast are
deterministic, we shall see in \S \ref{sec:convergence} that there is
some inherent noise in the computations; thus we conservatively set $c
= 1 \times 10^{-5}$ in our analysis.  

For the Gaussian process framework, sampled data are assumed to be
Normally distributed with means equal to the true function and
variance given by the sampling noise.  Moreover, a combination of any
subset of these points results in a Normal distribution.  Thus, we can
use the observed set of data, $\bA\subset \bP$, to predict the value
of $f$ for any $s_q \in \bP$.  This query point, $s_q$, will be Normally
distributed, ($N(\mu_{s_q}, \sigma_{s_q})$), with mean and variance given by
\begin{eqnarray}
\mu_{s_q} &=& \bar f_\bA + \Sigma_{\bA q}^T \Sigma_{\bA\bA}^{-1} (f_\bA
- \bar f_\bA) \label{k_mean}\\
\sigma^2_{s_q} &=& \Sigma_{\bA q}^T \Sigma_{\bA\bA}^{-1} \Sigma_{\bA q} \label{k_var}
\end{eqnarray}
where the elements of the matrix $\Sigma_{\bA\bA}$ and arrays
$\Sigma_{\bA q}$ and $f_\bA - \bar f_\bA$ are given by
\begin{eqnarray*}
\Sigma_{\bA \bA} [i,j] &=& \gamma(a_i, a_j)\\
\Sigma_{\bA q} [i] &=& \gamma(a_i, s_q)\\
(f_\bA - \bar f_\bA)[i] &=& f(s_i) - \bar f_\bA\\
\bar f_\bA &=& \frac{1}{|\bA|} \sum_{i=1}^{|\bA|} f(a_i)
\end{eqnarray*}
and the $a_i$'s and $a_j$'s are the observed data used to make an
inference: $a_i, a_j \in \bA$, $0\le i, j \le |\bA|$.


As given, for a set of $n$ observed points ($|\bA| = n$), prediction
with a Gaussian process requires
$O(n^3)$ time, as an $n \times n$ linear system of equations must be solved.
However, for many Gaussian process --- and ordinary kriging in particular
--- the correlation between two points decreases as a function of
distance.  Thus, the full Gaussian process model can be approximated well by a local
Gaussian process, where only the $k$ nearest neighbors of the query point are used
to compute the prediction value; this reduces the computation time to
$O(k^3+k\log(n))$ per prediction, since $O(k\log(n))$ time is required to find the
k-nearest neighbors using spatial indexing structures such as balanced
kd-trees.

\subsection{Algorithm} \label{sec:algorithm}
There are many well-known heuristics for computing where best
to perform the next experiment using a regression model, such as
that derived in \S \ref{model}.  Sampling strategies include picking the
point with the largest variance \citep{mackay1992,guestrin2005},
entropy or information gain. 

Sampling points based solely on variance is common in active learning
methods whose goal is to map out an entire function, as this will
minimize the expected error for prediction.  Moreover, the 
model variance predicted by local ordinary kriging 
is linear in the distance to the nearest neighbors.
As such, this strategy chooses points that are far from areas currently
searched, and thus will not get stuck in a specific location in
parameter space. However, this
strategy is known to over sample boundary regions \citep{mackay1992},
and ultimately samples the space evenly like a grid.
It is likely that large regions of the input space, $\mathcal{P}$,
fall well outside the confidence ball radius. In the 
progression of the algorithm, points in these regions may have large
variances but still not be within 2 or more standard deviations of the
boundary; these points are very unlikely to be near the confidence
ball radius.  Hence, a strategy that samples the entire space
evenly, using either a grid or a variance metric, can be extremely
inefficient for mapping function boundaries.

Information gain heuristics are also popular in the machine learning
community.  However in a continuous parameter space, computing the effect of adding
a new point is prohibitively expensive. Specifically, calculating the
information gain of a proposed sample requires integrating the 
difference between the current model and expected result of the
proposed sample over all space. Since our function approximator has
only local support for predictions, we can reduce this integral down
to the local region.  However on this local region, computing
the expected value of the model requires multiple matrix inversions
to account for differences in the 100 nearest neighbors over the local
region.   Even approximating this integral with a (small) finite sum,
was found to be prohibitively expensive.
Instead, we use a strategy that is a combination of entropy and
variance (both easy to compute), and is
related to information gain. For more discussion on sampling
strategies and their performance, we refer interested readers to
\cite{bryan2005}.

The method we use here, named ``Straddle'', combines the desire to
search the entire input space with that of refining our estimate
around known interesting regions.  We do this by picking points that
the model predicts are both close to the boundary and have large
variances using the following heuristic:
\[
\mathrm{straddle}(s_q) = 1.96  \sigma_{s_q} - \big|
\mu_{s_q} - t \big|.
\]
Note that the straddle heuristic chooses those points with large
variances which straddle the boundary.  In particular, if a point is
near the boundary, then $\mu_{s_q} \simeq t$ and 
this metric is equivalent to a variance-only metric, choosing
points that are distant from one another.
However, if the point is not on the boundary, then its score drops off
proportionally to the distance from the boundary.  The straddle score
for a point may be negative, which indicates that we predict that the
probability that the point is on a boundary is less that five
percent.  Note that the straddle algorithm scores points highest that
are both unknown and near the boundary, and thus gives scores that
intuitively are similar to that of information gain.

Our sampling strategy then consists of four steps.  First we model our
current knowledge using the Gaussian process described in \S
\ref{model}.  We then choose a set of candidate points randomly from the input
space and compute their mean and variances using the Gaussian process model.  Next,
we score these points using the Straddle heuristic, and 
select the highest scoring point.  Finally, we run the chosen point
through CMBFast and add use the result to refine our Gaussian process model.

Ideally, we  would like to analyze the
entire input space, and pick experiments in such a manner that
minimizes the number of experiments necessary.  However, as our
input space is infinite (the parameters are continuous), we need
a heuristic to quickly generate a large, but not unwieldy set of 
candidate points. 
\textit{A priori}, we have no information about the function we are trying to
model.  Therefore, in order to ensure that all
boundary segments of the true function are found (assuming sufficient
experimentation), it is necessary that candidate points be chosen such
that all infinitesimal hyper-rectangles in the input space have
non-zero probabilities of being chosen. 
We therefore choose candidate points uniformly at randomly from the
input space, as this satisfies the probability constraint and is
extremely quick.  We note that bad candidate points will be discarded
when their straddle scores are computed, and pose no problem for the
algorithm.

\section{Results} \label{sec:results}
Using the algorithm described in \S \ref{sec:algorithm}, we have
sampled just over 1.2 million CMBFast models creating a ``primary''
data set.  Additionally, we sampled another 100 thousand models
uniformly at random throughout the parameter space.
From the randomly sampled data, we find that less than
0.1\% of the parameter space searched is within the $2 \sigma$ confidence ball;
that is, our set of acceptable models (those within $2\sigma$) exclude
99.97\% of all possible models defined in Table \ref{paramtable}.
However, the
method we use to generate parameter vectors results in only 54\% of
the points being rejected by the hypothesis that the model and the
regressed fit are the same.  Thus, by actively searching through the
space, we are able to identify and efficiently map regions of interest, while
ignoring large areas of parameter space that result in models below
the $2\sigma$ level.  In \S \ref{sec:mcmc} we will see that our method
is much more data efficient than typical Bayesian methods.

\subsection{Confidence Interval Projections} \label{sec:intervals}

\begin{figure*}[!th]
 \begin{center}
 \plotone{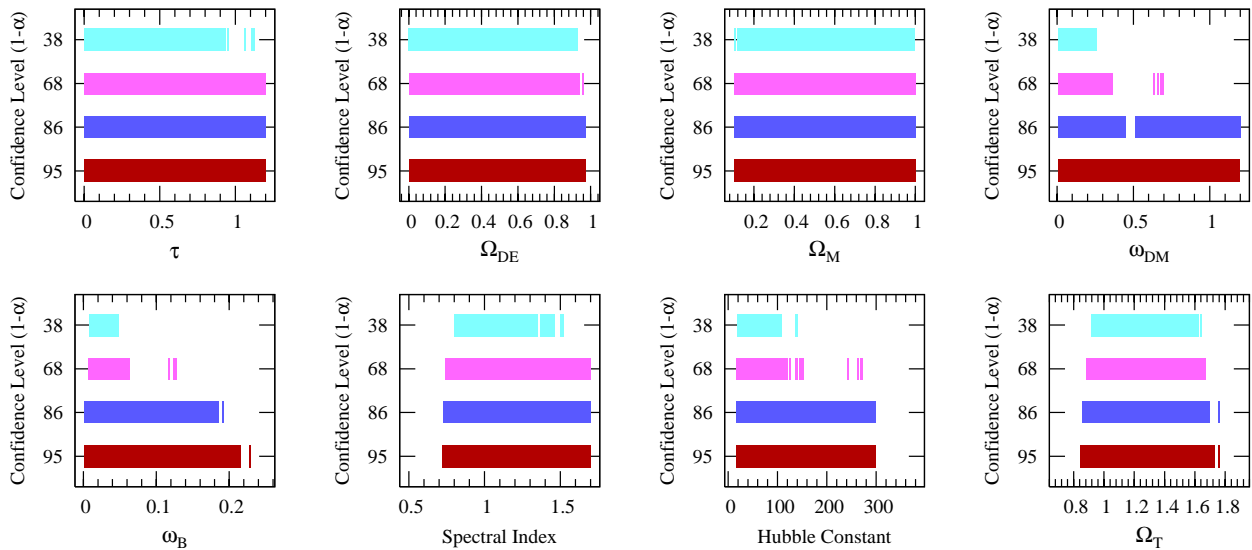}
 \end{center}
 \caption{Jointly valid confidence intervals for our cosmological
   parameters for four values of $1-\alpha$, corresponding to
   $\frac{1}{2} \sigma, \sigma, 1 \frac{1}{2} \sigma$ and $2\sigma$
   confidence levels, respectively.
   Areas of solid color indicate values for the given parameter
   that contain the true value of cosmological parameter with
   probability $1-\alpha$, regardless of the values of the remaining 6
   parameters.  
See the electronic edition of the Journal for a color version of this figure.}
\label{fig:results1d}
\end{figure*}

\begin{figure*}[p]
\begin{center}
\plotone{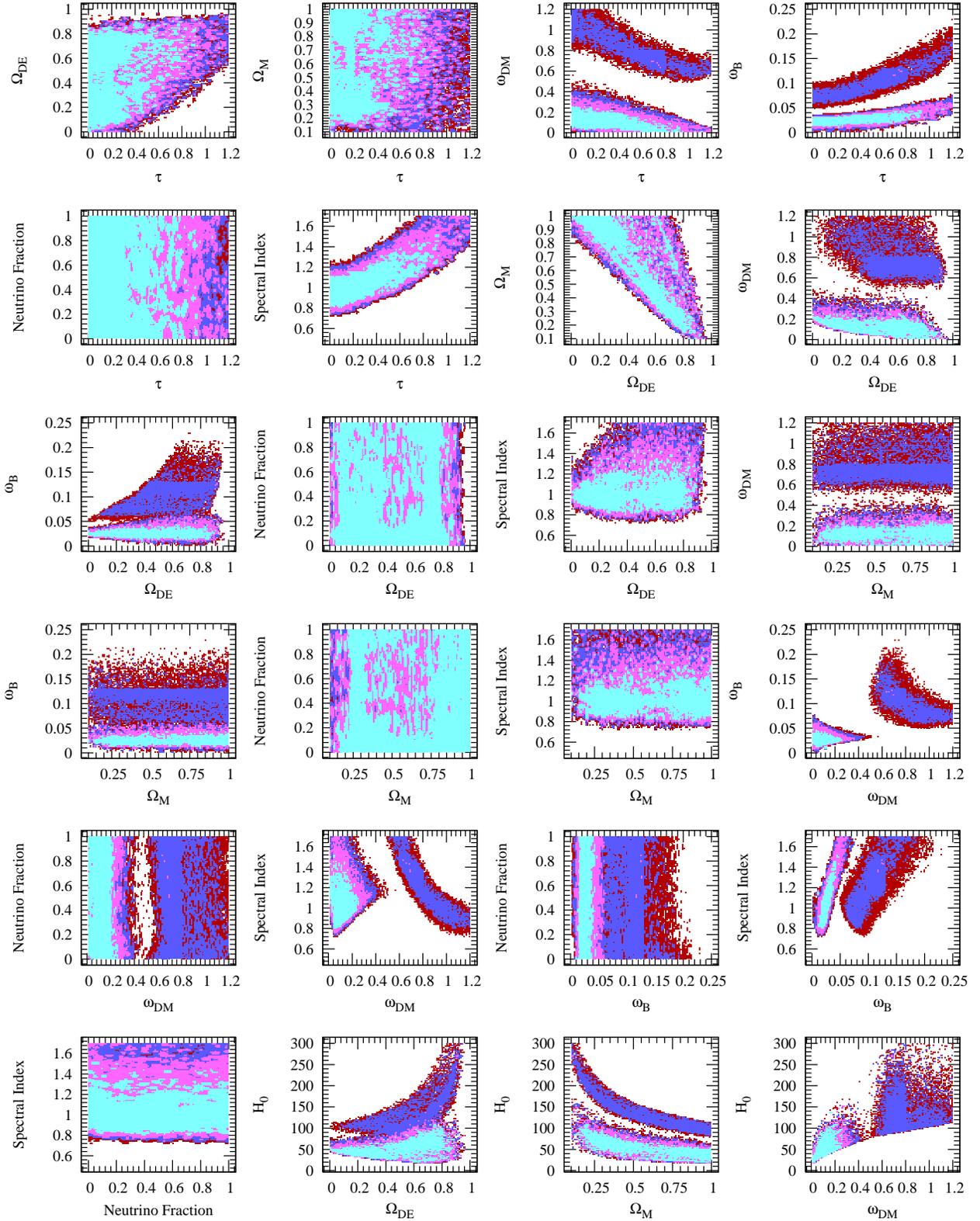}
\end{center}
\caption{Jointly valid confidence regions for pairs of cosmological
parameters, where the colors cyan, magenta, blue and red correspond to 
$\frac{1}{2} \sigma, \sigma, 1 \frac{1}{2} \sigma$ and $2\sigma$,
confidence levels respectively.
Areas of solid color indicate values for the given
pair of fixed (plotted) parameters that contain the true value of 
cosmological parameter with probability $1-\alpha$, regardless of the
values of the remaining 5  parameters.
Note there are two disjoint regions in parameter space
   which are above the $2\sigma$ confidence interval.
See the electronic edition of the Journal for a color version of this figure.}
\label{fig:results2d}
\end{figure*}

\begin{figure*}[t]
\begin{center}
\plotone{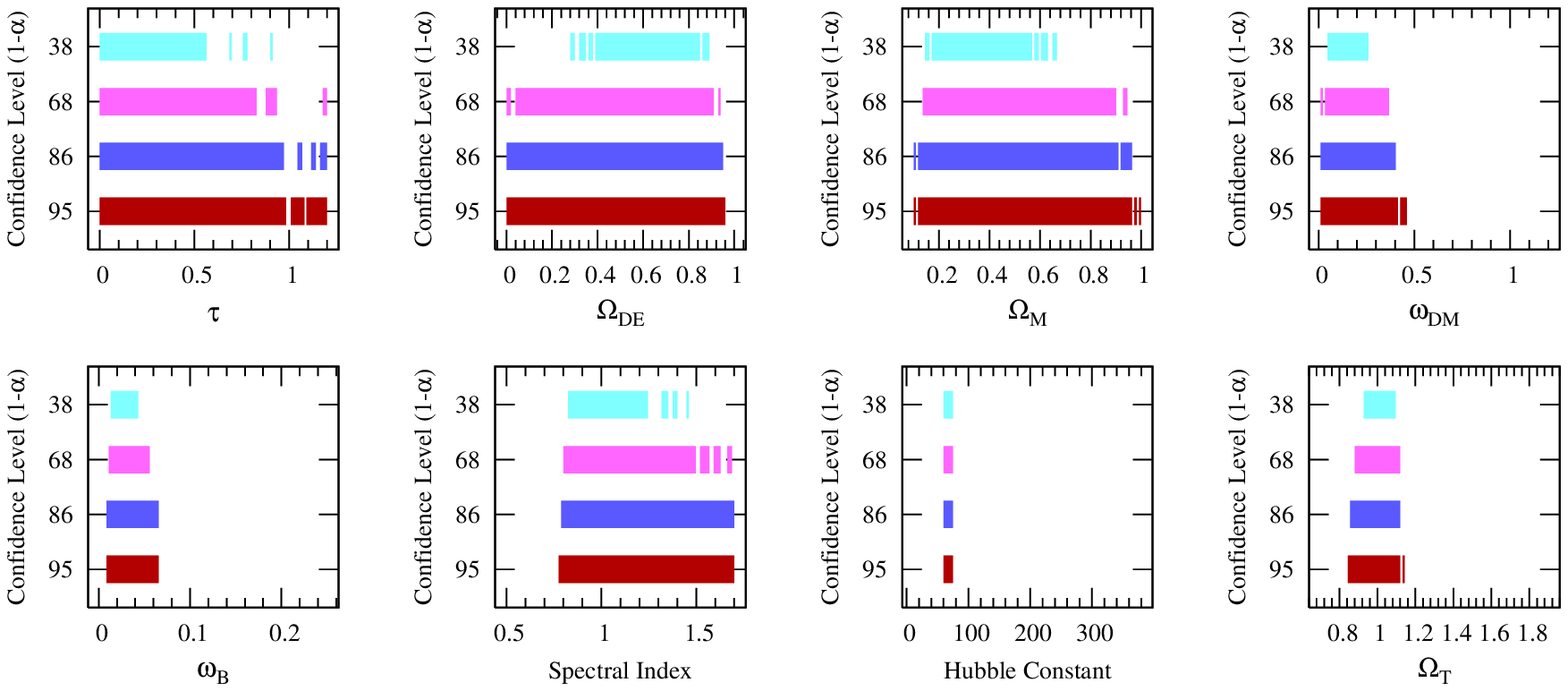}
\end{center}
\caption{Jointly valid confidence intervals for our cosmological
parameters, where we assume that that the value of $H_0$ is between 60
and $75 \mpc$.  Areas of solid color
indicate values for the given parameter that contain the true value of
cosmological parameter with probability $1-\alpha$, regardless of the
values of the remaining 6 parameters.  See the electronic edition of
the Journal for a color version of this figure.}
\label{fig:results1d:h0}
\end{figure*}

\begin{figure*}[p]
\begin{center}
\plotone{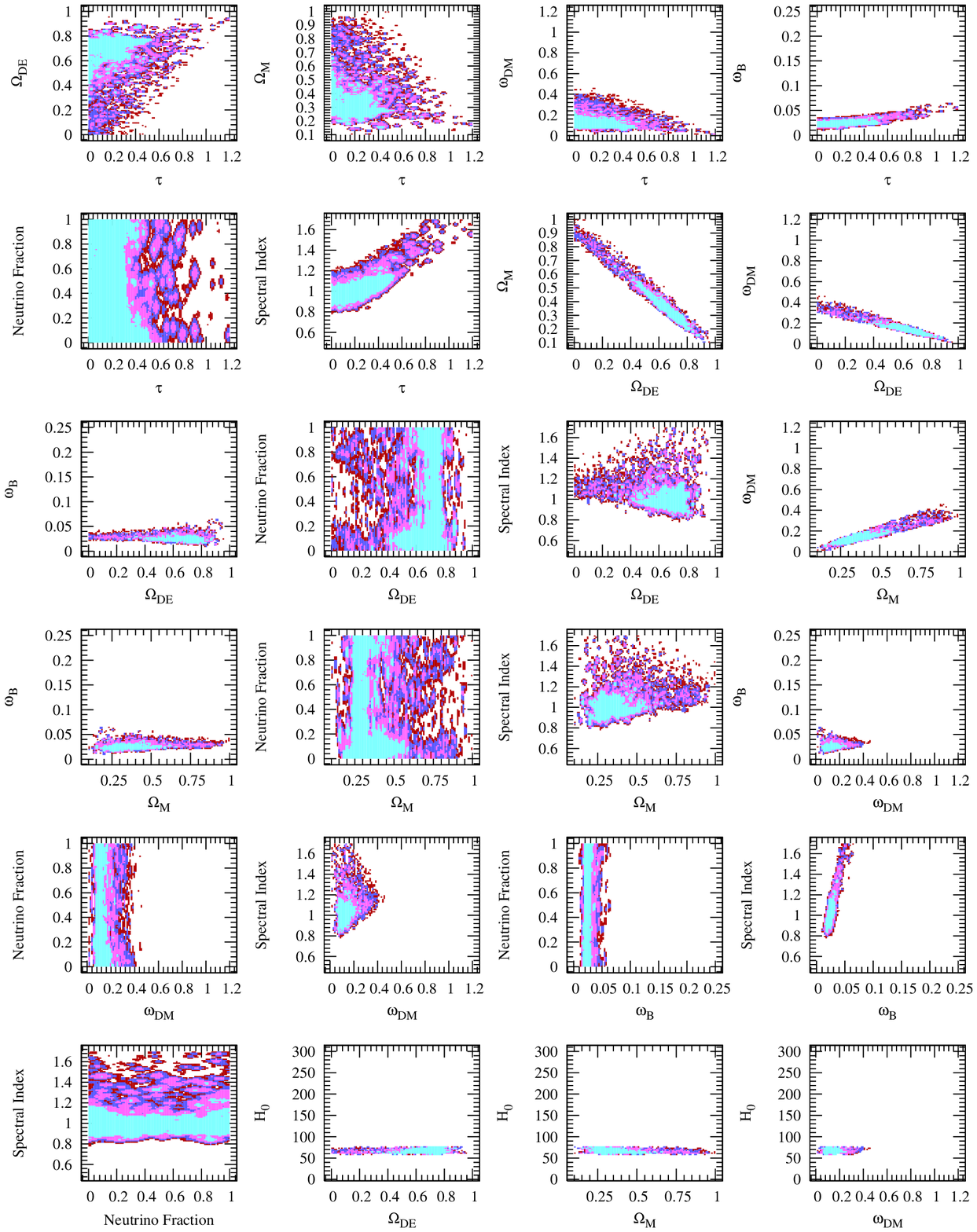}
\end{center}
\caption{Jointly valid confidence regions for pairs of cosmological
parameters, where we assume that that the value of  $H_0$ is between
60 and $75 \mpc$. The colors
cyan, magenta, blue and red correspond to 
$\frac{1}{2} \sigma, \sigma, 1 \frac{1}{2} \sigma$ and $2\sigma$,
confidence levels, respectively.
Areas of solid color indicate values for the given
pair of fixed (plotted) parameters that contain the true value of 
cosmological parameter with probability $1-\alpha$, 
regardless of the values of the remaining 5  parameters.
Note that the constraint on $H_0$ eliminates the secondary confidence 
region found in Figure \ref{fig:results2d}.
See the electronic edition of the Journal for a color version of this figure.}
\label{fig:results2d:h0}
\end{figure*}


The result of running the 1.2 million models contained in the primary
data set is a set of
disjoint, seven dimensional ``confidence regions'' in parameter space
which contain all models that fall within our $1-\alpha$ confidence
ball.  In each of these regions, the confidence interval for a
particular parameter is given by the range of values that parameter
takes in that region. Thus, the confidence interval for a particular
parameter will be a function of which sets of regions we consider.  

If we put no restrictions on the values of the other 6 parameters,
then the confidence interval of a parameter will be the union of
the confidence intervals for that parameter for all confidence
regions. We plot these unrestricted confidence intervals in Figure
\ref{fig:results1d} for four values of $1-\alpha$.  
Intuitively, Figure \ref{fig:results1d} can be interpreted as stating
that for any value of a parameter that lies within the depicted
$1-\alpha$ confidence interval, there exists at least one
combination of the remaining six parameters such that the resulting
parameter vector lies within one of the $1-\alpha$ confidence regions.  

In Figure \ref{fig:results2d} we depict results of interactions between pairs
of parameters on the computed confidence regions.  As with the 1D
projections in Figure \ref{fig:results1d}, points in Figure
\ref{fig:results2d} which are denoted to be within the $1-\alpha$
confidence ball, are points where given the particular values of the
two fixed cosmological parameters --- those being explicitly plotted
on the $x$ and $y$ axes, --- there exists some values for the other 5
parameters such that the resulting parameter vector is within the
$1-\alpha$ confidence region.   While some plots show that most
combinations of the fixed parameters are within the 95\% confidence
ball providing minimal constraints on parameters describing the
Universe, others, such as $\omega_\mathrm{DM}$ versus
$\omega_\mathrm{B}$ (4\ith row, 4\ith column), show strong
constraints. 

Areas in Figure \ref{fig:results2d} which are blank (white), are areas that are
rejected at the 95\% confidence level; for these combinations of fixed
parameters, there exists no combination of the other five parameters,
such that the resulting vector is within any of our confidence regions.
In particular, the plot of $\Omega_\mathrm{DE}$ versus
$\Omega_\mathrm{M}$ (2\ind row, 3\ird column) illustrates that
$\Omega_\mathrm{Total} \gtrsim 0.9$, while the plot of $\omega_\mathrm{DM}$
versus $\omega_\mathrm{B}$  shows that there are at least two disjoint
confidence regions in our seven dimensional space.  These disjoint
regions in Figure \ref{fig:results2d} correspond directly to the split
confidence intervals observed in Figure \ref{fig:results1d}.

The disjoint regions observed in Figure \ref{fig:results2d}, such as
the plot of $\omega_\mathrm{DM}$ vs. $\omega_\mathrm{B}$, indicate
that there are at least two disjoint confidence regions in the
parameter space. 
These disjoint regions can also be seen in the 1D projections of
$\omega_\mathrm{DM}$, $\omega_\mathrm{B}$, and $H_0$ shown in Figure \ref{fig:results1d}.
We defer further discussion of the disjoint confidence regions 
to \S \ref{sec:connectivity}. Smaller splits in the confidence
intervals observed in nearly 
every plot in Figure \ref{fig:results1d} are a result of the fact that
CMBFast does not return models which are perfectly continuous in the
parameter space.  While one may expect the derived confidence level to
be smooth in parameter space, this is not the case.
We observe small discretizations and
inconsistencies in the power spectrum model, which result in the
confidence ball having a jagged, nebulous surface (as observed in
Figure \ref{fig:results2d}), rather than a perfectly smooth one.  We will
elaborate on this observation in \S \ref{sec:convergence}. 

As illustrated in Figure \ref{fig:results1d}, the confidence intervals
for most parameters are not well constrained by the WMAP data  alone.
In particular, the constraint
on the Hubble constant, $H_0$, is so weak as to allow values between
15 and 300 at the two sigma level; even at the one sigma level, $H_0$
ranges between $15$ and $150$ with additional fits at $H_0 \sim 250$.
The confidence intervals derived here cover the Bayesian credible
intervals found in the literature using a
variety of techniques (e.g. \cite{tegmark2001, spergel2003,
  spergel2006}), as shown in Table \ref{tab:compare}.
While the results in Table \ref{tab:compare} are 
approximately centered on the same values,
we are not in any way attempting to argue that the allowed parameter
ranges are better, or worse, than those derived from alternative methods,
as the comparison of credible (Bayesian) vs. valid (frequentist)
parameter ranges is non-trivial and outside the scope of this work.
A discussion of difference between the Bayesian and frequentist 
interpretations is given in \S \ref{sec:bayesvsfreq}.

\begin{table*}
\begin{center}
{\footnotesize
\begin{tabular}{c r@{ - }l r@{ - }l r@{ - }l | r@{ - }l r@{ - }l}
\hline
& 
\multicolumn{2}{c}{No} &
\multicolumn{2}{c}{} &
\multicolumn{2}{c|}{$n_s < 1$} &
\multicolumn{2}{c}{Spergel} &
\multicolumn{2}{c}{Spergel}\\
Parameter &
\multicolumn{2}{c}{Constraints} &
\multicolumn{2}{c}{$ 60 \le H_0 \le 75$} &
\multicolumn{2}{c|}{$ 60 \le H_0 \le 75$} &
\multicolumn{2}{c}{et al. (2003)} &
\multicolumn{2}{c}{et al. (2006)}\\
\hline
\hline
$\tau$ 
& 0 & 1.2 
& \multicolumn{2}{c}{0 - 0.94, 1.17 - 1.2} 
& 0 & 0.4 
& 0.095  & 0.242 
& 0.058 & 0.117 
\\
$\Omega_\mathrm{DE}$ 
& 0 & 0.94   
& 0 & 0.94   
& 0.39 & 0.9 
& \multicolumn{2}{c}{} 
& \multicolumn{2}{c}{} 
\\
$\Omega_\mathrm{M}$ 
& 0 & 1.0   
& 0.13 & 0.95 
& 0.13 & 0.59 
& 0.22  & 0.36  
& 0.199 & 0.273 
\\
$\omega_{\mathrm{DM}}$ 
& \multicolumn{2}{c}{0 - 0.36, 0.62 - 0.70} 
& 0.0 & 0.36 
& 0.03 & 0.2 
& \multicolumn{2}{c}{} 
& \multicolumn{2}{c}{} 
\\
$100\omega_{\mathrm{B}}$ 
& \multicolumn{2}{c}{0.5 - 6.2, 11.5 - 12.7} 
& 1.3   & 5.5 
& 1.3   & 3.2 
& 2.26  & 2.51  
& 2.15  & 2.31  
\\
$f_\nu$ 
& 0 & 1 
& 0 & 1 
& 0 & 1 
& \multicolumn{2}{c}{} 
& \multicolumn{2}{c}{} 
\\
$n_s$ 
& 0.73 & 1.7 
& 0.8  & 1.7 
& 0.84 & \textit{1.0} 
& 0.95  & 1.03   
& 0.944 & 0.978  
\\
$\sigma_8$ 
& \multicolumn{2}{c}{} 
& \multicolumn{2}{c}{} 
& \multicolumn{2}{c|}{} 
& 0.82  & 1.02   
& 0.71  & 0.81   
\\
$H_0$ 
& \multicolumn{2}{c}{17 - 135, 243 - 272} 
& \textit{60} & \textit{75} 
& \textit{60} & \textit{75} 
& 67    & 77     
& 70.3  & 76.7   
\\
\hline
\end{tabular}}
\end{center}
\caption{Derived 68\% confidence intervals.  Those to the left of the solid
line are derived from Figures \ref{fig:results1d},
\ref{fig:results1d:h0} and \ref{fig:results1d:nsh0} respectively,
while those to the right are quoted from referenced literature.}
\label{tab:compare}
\end{table*}

While this assessment may appear bleak, there is 
underlying structure to the confidence regions, hinted at by the
disjoint regions in Figure \ref{fig:results2d}.   Suppose we restrict
the range of a subset of our parameters and then compute the
confidence intervals for the remaining parameters.
Since our statistical model is independent of the ranges searched, we can
compute these conditional confidence intervals without re-running any
models.  For any restriction of our parameter space, 
the confidence interval for a parameter of interest will
be the union of the confidence intervals for that parameter over those
confidence regions which obey our restriction.  For example, in Figures
\ref{fig:results1d:h0} and \ref{fig:results2d:h0} we show the effect
on the confidence intervals and regions, respectively,
of imposing the restriction that $H_0$  is between
$60$ and $75\mpc$.   Note that with this
restriction on $H_0$, the confidence intervals agree much better with the
current estimate of the cosmological matter/energy budget and strongly
suggest that $\Omega_\mathrm{Total} = 1$.

This analysis exhibits the power of our statistical
inference technique: we can test constraints on one parameter,
and see their effects on the remaining parameters without additional
CMBFast computation or invalidation of statistical inferences.  To
this end, we have created a graphical  interface that can be used to
apply constraints and view the resulting effects in real time; this
tool, along with the necessary data files, can be downloaded from
\url{http://gs3636.sp.cs.cmu.edu/visualizer/}.  

In the Bayesian view, the tightening of the allowable regions between
Figures \ref{fig:results1d} and  \ref{fig:results1d:h0}
and Figures \ref{fig:results2d} and \ref{fig:results2d:h0}
is analogous to what would occur when priors
(either informative or non-informative) are applied. Such
priors are  universally applied in CMB cosmological analyses.
As an example of how we can use this technique to better
understand the cosmological confidence surface, we focus
in on one or two parameters and utilize the graphical interface
described above.

\begin{figure*}[t]
\begin{center}
\plotone{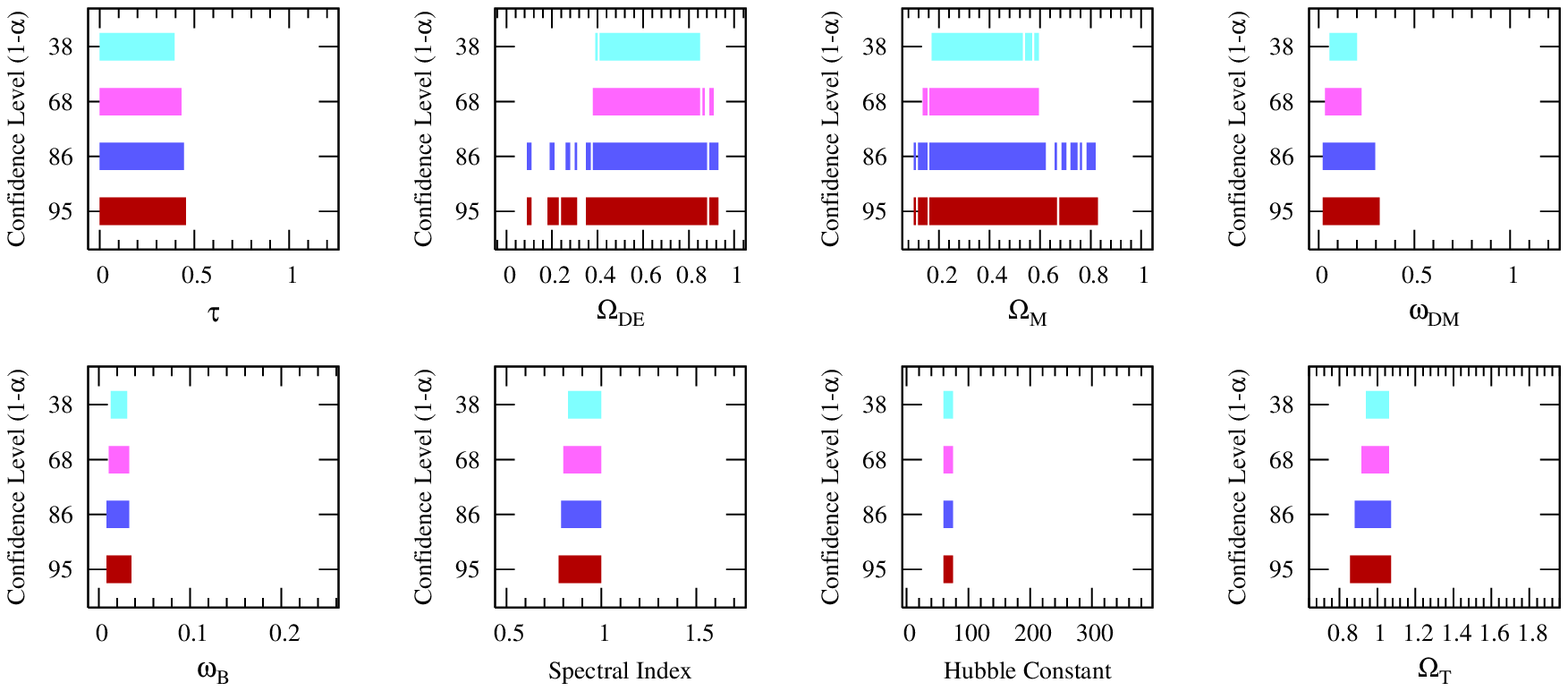}
\end{center}
\caption{Jointly valid confidence intervals for our cosmological
parameters, where we assume that $60
\mpc \le H_0 \le 75 \mpc$ and $n_s < 1$.  Areas of solid
color indicate values for the given parameter that contain the true
value of cosmological parameter with probability  $1-\alpha$,
regardless of the values of the remaining 6 parameters.
See the electronic edition of the Journal for a color
   version of this figure.}
\label{fig:results1d:nsh0}
\end{figure*}

WMAP Three Year data show that a scale invariant spectra
($n_s = 1$) is not a good fit to the WMAP Three Year data alone.
If we place both the constraint that $n_s < 1$ and that 
$ 60 \mpc \le H_0 \le 75\mpc$ on the WMAP One Year data, we see
in Figure \ref{fig:results1d:nsh0} that $\tau, \omega_\mathrm{B}$, and 
$\omega_\mathrm{DM}$ are much better constrained. More importantly, we
see that the allowable ranges on $\omega_\mathrm{DM}$ are forced into a single
confidence range, in agreement with previous studies \cite{spergel2003}.

Exploring the high $\omega_\mathrm{DM}$ space shown in Figure
\ref{fig:results1d}, we find that models consistent with
high $\omega_\mathrm{DM}$ have large values of $\omega_\mathrm{B}$ ($> 0.05$),
as well as large Hubble constants ($>100\mpc$). Both of these parameters are
much better constrained in the WMAP Three Year data. This
leads us to predict that the second confidence surface peak in
the WMAP Three Year Data is less significant than in the
WMAP One Year data (although this has yet to be shown).

%

%
%
%
%
%
%

\subsection{Convergence} \label{sec:convergence}
Ideally, one would like to prove that our mapping from confidence
ball radius to parameter space has converged.  This could be done, for
instance, by proving that our approximating model of spectrum distance
as a function of cosmological parameters -- that is our Gaussian
process -- has converged to the true values in those areas where the
true values are near the radius of the $1-\alpha$ confidence ball.
However, this effort has been confounded by a lack of continuity in
the results returned by CMBFast.  The method presented in this paper
is not more susceptible to discontinuities than other techniques.
Indeed, the convergence of most, if not all, inference methods will be
adversely effected by the discontinuities of CMBFast models we observe
in parameter space.

\begin{figure}
\begin{center}
\noindent
\plotone{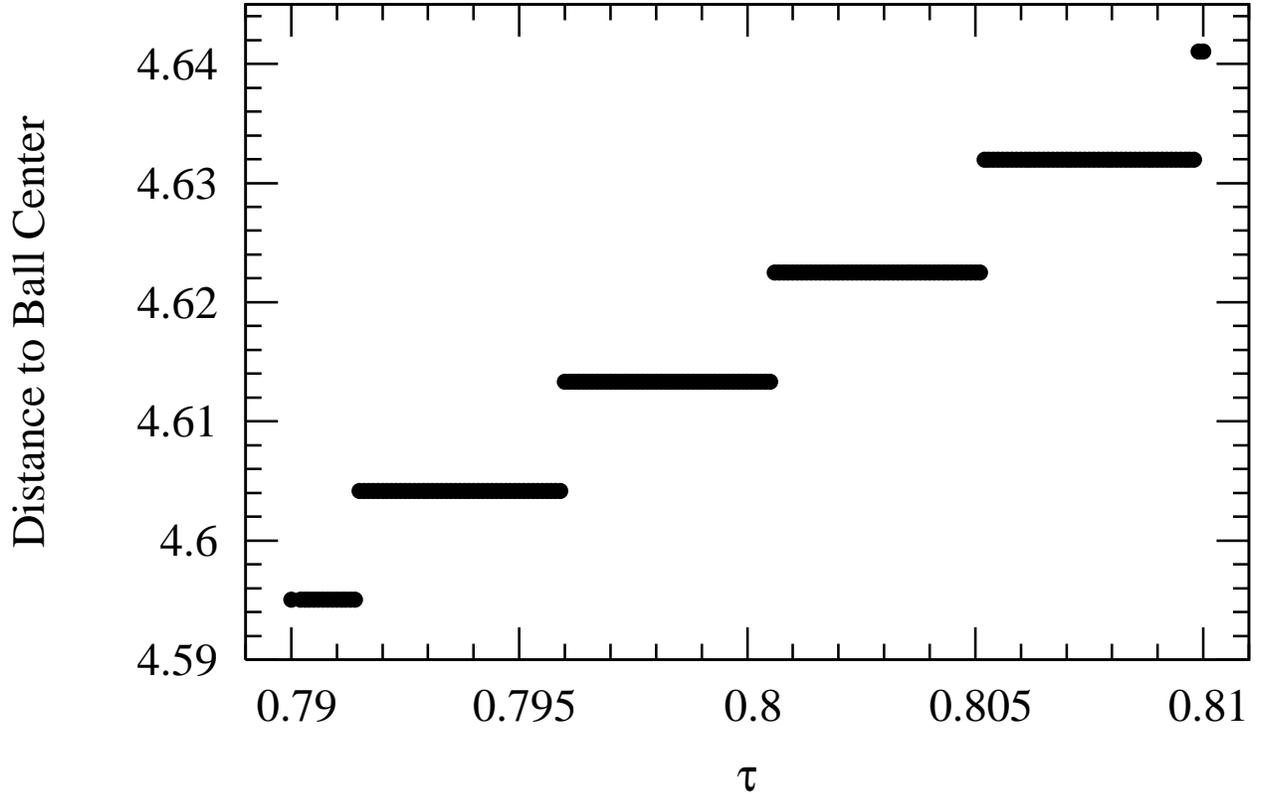}
\end{center}
\caption{A plot of spectra distance as a function of $\tau$, with
 all other parameters fixed, showing the discretization of CMBFast.
For these experiments 
$\vec x = \{\tau, \Omega_\mathrm{DE}, \Omega_\mathrm{M},
 \omega_\mathrm{DM}, \omega_\mathrm{B}, f_\nu, n_s\}$ $= 
\{\tau, 0.0, 0.2, 0.8, 0.003, 0.0, 1.2\}$.}
\label{fig:smooth}
\end{figure}

One standard assumption of function approximators is that of
smoothness; that is that the underlying function to be modeled is
continuous and differentiable.  For Gaussian processes, this
assumption motivates the usage of a covariance matrix in determining
the relative weights of known samples when estimating values for unknown points.  In
this paper, we have also assumed that the covariance function is fixed over
the entire space -- that is that the underlying covariance is
isotropic and homogeneous.  These assumptions allow us to compute
error bounds for each point in space, and enable us to determine when
the model has converged to the underlying function.  

\begin{figure*}
\begin{center}
\noindent
\plotone{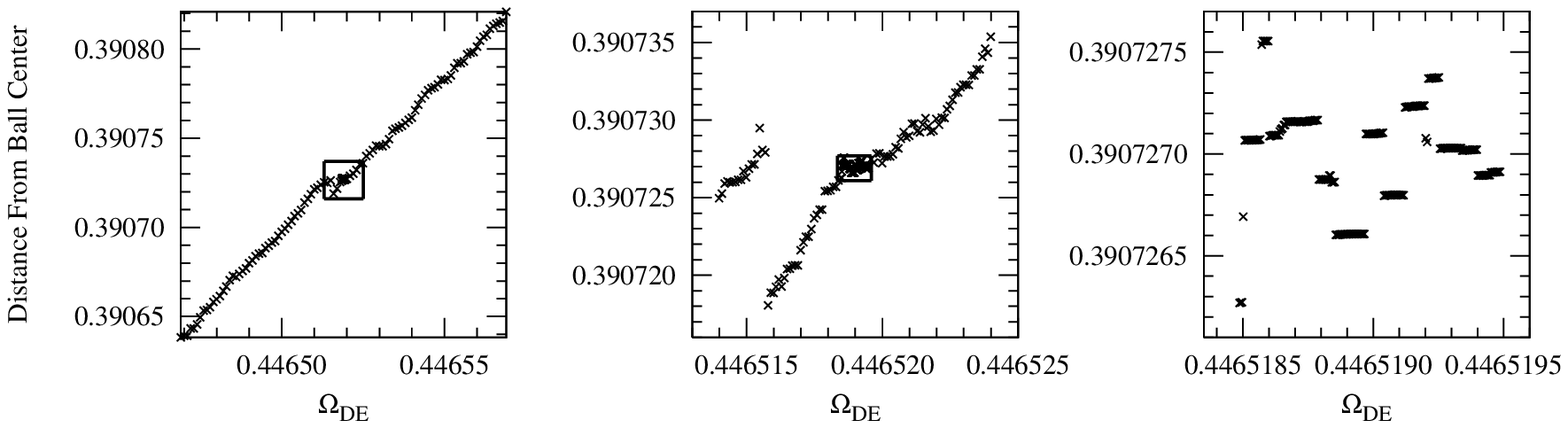}
\end{center}
\caption{A plot of spectra distance as a function of
$\Omega_\mathrm{DE}$, with all other parameters fixed.  The square 
boxes in each of the left two plots denotes the area enlarged in the
neighboring plot to the right. Note that while on the global scales,
(A), the mapping appears to be smooth, closer inspection (B),(C)
reveal numerical errors resulting from approximations used in CMBFast.} 
\label{fig:smooth2}
\end{figure*}

However, experimentation shows that the underlying CMBFast function
does not fulfill the continuous and differentiable assumptions, as
shown in Figures \ref{fig:smooth} and \ref{fig:smooth2}.
Both figures were produced by plotting the resulting model distance
as we varied one parameter and kept the other six parameters fixed.  Figure
\ref{fig:smooth} shows a discretization effect that we believe is a
result of integral approximations done by CMBFast.  Discretization
effects are common in simulated environments and it is reasonable to
assume that the true function varies smoothly.  More startling are the
discontinuities revealed in Figure \ref{fig:smooth2}.  Figure \ref{fig:smooth2} shows
that while on a broad scale the CMBFast function appears smooth, when
one looks closer and closer, the function begins to act quite
erratically.  Of particular interest are the large discontinuity at
$\Omega_\mathrm{DE} = 0.446516$ and the seemingly random deviations
from a smooth function throughout the entire range.  These fluctuations 
in distance are not caused by random noise from CMBFast; 
CMBFast's output is deterministic given an input parameter
vector.   

There are two important implications of the results in Figures
\ref{fig:smooth} and \ref{fig:smooth2}.  First, we note that 
when parameter values result in spectra that are very close to the
confidence ball radius, it is impossible to predict which side of
the boundary a given point will be on, due to the inherent noise in
CMBFast.  For regions where many points are near the confidence ball
radius, we will obtain spotty, jagged boundaries between those areas
in the ball and those not.
Secondly, the effects plotted in
Figures \ref{fig:smooth} and \ref{fig:smooth2} do not appear on the
same range scales. This makes it more difficult to determine the
correct level of smoothing, and hence discover the true underlying
function.  Thus, while it is still possible to deduce approximate
covariances among the variables, it becomes impossible to ensure the
model has correctly converged to the true model.

We note that this lack of continuity will adversely effect the
convergence of any model that relies on the smoothness of the
underlying function, be it MCMC or Gaussian processes.  
In the case of MCMC, 
the discontinuities in the variance weighted sum of squares between
the models computed by CMBFast and the data require that comprehensive
sampling of the posterior be performed to ensure that the peaks and
valleys in any local region are correctly averaged out, ensuring that
the integral over the posterior is correctly computed.  While we
can run both methods in a mode that smooths over these
discontinuities (by effectively ignoring them), we must realize that
the resulting algorithms will converge to a solution that is
incorrect.  Additionally, increasing the sampling of either
algorithm would eventually turn up the existence of these
discontinuities, and the system would jump from an apparent
convergence in the smoothed case, to a new convergence where
discontinuities are considered.  We elaborate on this idea further in
\S \ref{sec:mcmc}.

\subsection{Connectivity} \label{sec:connectivity}
As Figure \ref{fig:results2d} shows, there are two main peaks that lie above
the $1\sigma$ confidence ball radius.  As a test of the 
function approximator's convergence, we conducted focused tests to see if these
peaks were truly connected.  In particular, we used the semi-variance
matrix of the Gaussian process to compute the maximal influence
distance from a given point one could travel before possibly
encountering the $1-\alpha$ confidence ball radius.  We then created
clusters of points above the 68\% confidence ball radius using a
friends-of-friends algorithm; that is, a point is added to an existing
group if it is within the maximal influence distance of any point
currently in the group.  Starting with all points in their own groups,
we first passed through the data, merging groups where possible.
Then, additional points were sampled between existing groups, using an
A$^*$ like algorithm \citep{hart1968}.  For two groups $A$ and $B$, we
found the point, $x$, in $A$ that was closest to any point in $B$.  We
then created a set of candidate points within the influence distance
of $x$, and add them to a queue, $\bQ$, sorted according to their
distances to $B$.  We then take the point $p$ from $\bQ$ that is closest
to $B$ run it through CMBFast and compare to our confidence ball.  If
$p$ is within our confidence radius, then we create 
candidate points for $p$ (just as we did for $x$) and add them to 
$\bQ$.  Otherwise, we remove $p$ from $\bQ$.
This procedure is repeated until either $B$ is within the influence
distance of $p$ or we exhaust $\bQ$.  

The primary data set contained roughly 2000 distinct groups, which were
quickly merged using the friends-of-friends algorithm.  This left us with
2 major clusters shown in Figure \ref{fig:results2d}.  
Using the algorithm noted above,
we were unable to find connections between the main peak and the
secondary peak, even after multiple attempts starting from
different locations.  We believe that there exists no 
smooth transition of variable parameters that leads from the
concordance to the secondary peak.  The second peak is not just an
extension of the concordance peak that appears disjoint due to under
sampling or projection effects.

\section{Comparison to Alternative Methods of Statistical
  Inference} \label{sec:comparison}

In \S \ref{sec:results}, we showed that the results of our
technique are quite similar to other statistical inference methods
currently employed in the literature.  Let us now relate our method
to other inference techniques, and point out a few subtle, but
remarkable, distinctions between them.

\subsection{$\chi^2$ Tests}
The method presented in \S \ref{sec:nonparametric} can be
succinctly described as a method which computes the weighted sum of
squares of the regressed fit and the test spectrum at the data
points and rejects the hypothesis that the test spectrum could be
generated by the data if the weighted sum is greater than the constant
given in Equation \ref{conf0}.
Intuitively, this process is quite similar to using a $\chi^2$ test, 
with two important differences.  

First, our technique is centered
around a nonparametric fit to the data, not the data themselves.  As a
result, our method is approximately centered on the true underlying
function, $f$, as opposed to the noisy observations of $f$.
The
implication is that our method is less affected by noise in the data,
than simple $\chi^2$ tests.
In particular, we have observed that $\chi^2$ tests will reject all
models in cases where there is a single outlier $4\sigma$ from the maximum
likelihood estimate fit.  By initially fitting a nonparametric
function to the data and then using this function to compute
sum-of-squares distances, we are much less susceptible errors
caused by noisy outliers. 

\begin{figure}
\begin{center}
\plotone{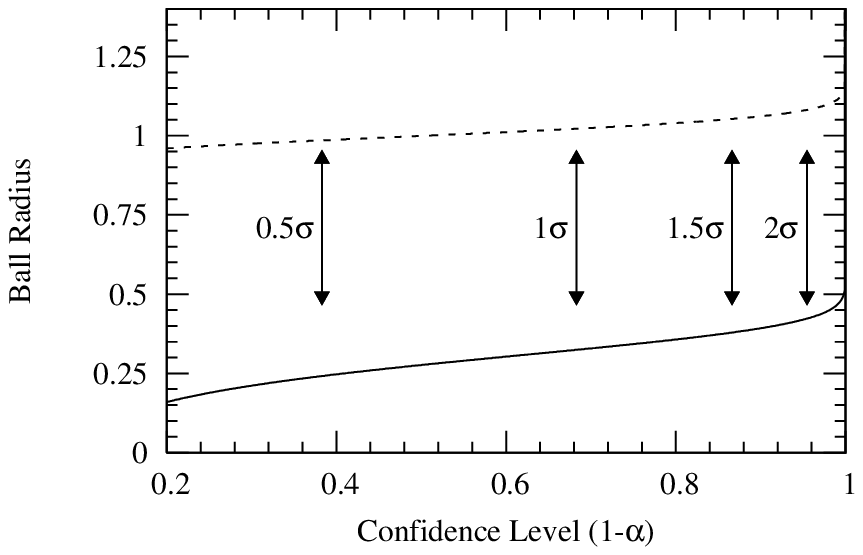}
\end{center}
\caption{Radius of our non-parametric confidence ball as a function of
  confidence level (solid).  The reduced $\chi^2$ ball is shown for
  comparison (dashed). Arrows depict $\frac{1}{2}, 1, 1\frac{1}{2}$ and
  $2\sigma$ respectively.}
\label{fig:distance_alpha}
\end{figure}

Secondly, the radius computed using the pivot process is smaller than
the $\chi^2$ radius, as we consider the Gaussian errors of all points
as an ensemble, not individually as with $\chi^2$ tests.  The smaller
radius of the pivot process translates directly into smaller confidence regions
as compared with those found using $\chi^2$ tests.  This allows
us to reject more of the hypothesis test models, and subsequently return tighter
bounds on the parameters of interest. The confidence ball test has
more statistical power than does the $\chi^2$ test.  A comparison of
the relative widths of the confidence and $\chi^2$ balls is shown in 
Figure \ref{fig:distance_alpha}.   

\subsection{Bayesian Techniques} \label{sec:mcmc}
As noted in \S \ref{sec:introduction}, most CMB power spectrum
parameter estimations to date have been done via 
Bayesian techniques (e.g., \cite{knox2001, gupta2002, spergel2003,
jimenez2004, dunkley2005}).  Since the prior distribution is not
conjugate on the likelihood, computing the posterior involves
estimating an integral over the entire space spanned by the prior. 
Perhaps the most straight-forward way to compute this integral is
with an evenly-spaced grid with $n$ points per parameter.  For this
approach, one pre-specifies a $d$-dimensional grid (where $d$ is the
number of parameters of interest) and computes the posterior at the
center of each grid cell.  The integral is then (approximately) the
sum of the posterior at each grid cell, and the $1 - \alpha$ credible
intervals can be determined (usually by marginalization) to be the
smallest range for a given parameter that contains $1-\alpha$ of the
posterior probability.   While straight forward, this approach scales
exponentially with dimension, and hence is infeasible for even moderate
dimensions; we estimate that a grid based approach, using CMBFast and
seven parameters (similar to our method), with just 10 grid
spacings per parameter would take over 100 years on a single computer.

As a result of the dimensionality problem, Markov Chain Monte Carlo
(MCMC) has become an increasingly popular approach for
estimating posteriors due to their (perceived) computational
efficiency (e.g \cite{gupta2002, jimenez2004, sandvik2004,
dunkley2005,chu2005}).
In the MCMC technique, new samples are often derived using the
Metropolis-Hastings algorithm.  The Metropolis-Hastings algorithm
chooses a new sample $x$ from some arbitrary (pre-specified) proposal
distribution defined over the  $d$-dimensional parameter space based on the
previous sample and then accepts or rejects $x$
based on the ratio of the proposed and current posterior density (when
the proposal distribution is symmetric, as is common).  
The algorithm samples the input space roughly in  
proportion to the expected probability of each location.  


Theoretically MCMC using Metropolis-Hastings algorithm
converges almost surely to the stationary distribution (the
posterior) in the limit of infinite sampling.  However, it is quite
difficult to determine if convergence has been met with a finite number
samples.  In particular, if a posterior is comprised 
by two narrow, spatially separated Gaussians, then the probability of
transition from one Gaussian to the other will be vanishingly small.
Thus, after the chain has rattled around in one of the peaks for a
while, it will appear that the chain has converged; however, after
some finite amount of time, the chain will suddenly jump to the other
peak, revealing that the initial indications of convergence were
incorrect.  As this example illustrates, if the Markov chain is run
with too few examples, the resulting credible intervals will be too
narrow, and thus will not truly contain $1-\alpha$ of the probability
mass. Thus, the consequence of lack of true convergence is artificially
small credible intervals.  This problem is usually skirted by assuming
that there are no small isolated peaks, computing multiple independent
chains and comparing the results to illustrate convergence.
Additionally, \cite{dunkley2005} and others have proposed alternative
methods to detect convergence.  However, none of these methods are able to
prove convergence with a limited number of CMBFast runs.

Moreover, as we noted in \S \ref{sec:introduction}, MCMC is designed
to draw samples from an unknown distribution, not to search that distribution.
As a result, MCMC algorithms explicitly spend a large number of samples
on high-likelihood regions, and a minimal number on low-likelihood
regions.  However, when we are computing $1-\alpha$ confidence
intervals, it is the low-likelihood regions (those around the
$1-\alpha$ boundary) that we are interested in.  In contrast, a search
algorithm that can directly look up the likelihood of a sample
has no reason to spend a large number of samples near the peak of the
distribution, and can instead focus on the boundary in question.

\begin{figure*}
\begin{center}
\plotone{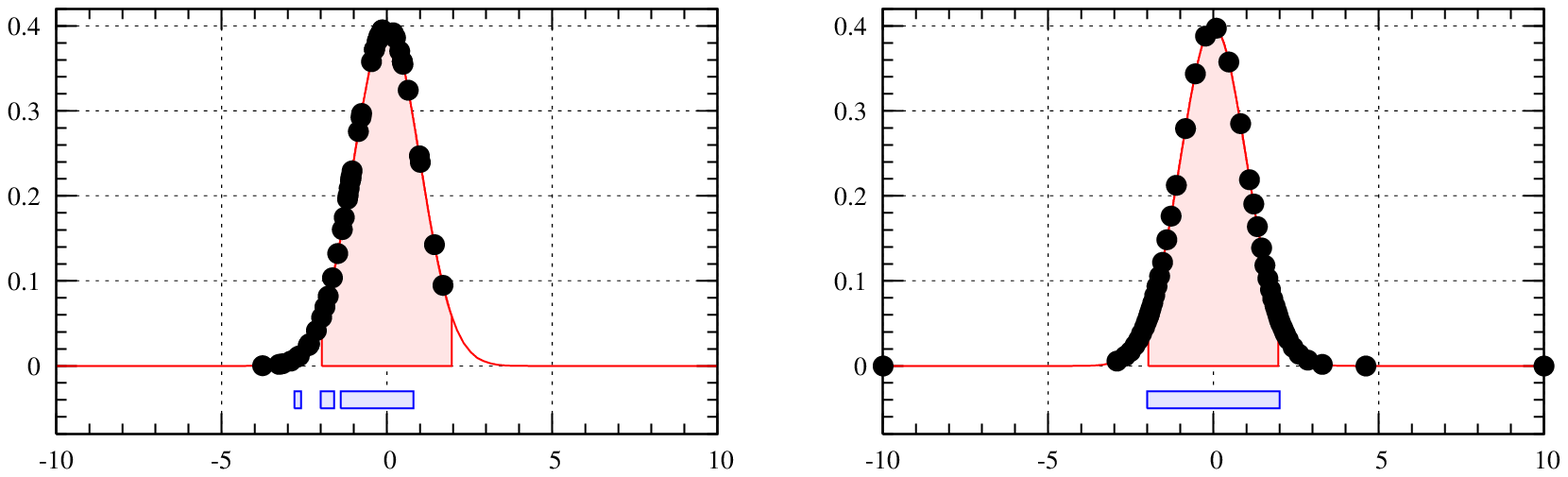}
\end{center}
\caption{Distribution of experiments run by MCMC (left) and our
  algorithm (right).  Black dots denote 50 experiments run in
  order to determine 
  the 95\% credible / confidence interval (shaded red area) for a
  standard normal 
  distribution (solid red line).  Shaded blue areas below the normal
  curves indicate the credible / confidence intervals derived for the
  50 samples chosen.  See the electronic edition of the Journal for a
  color version of this 
figure.}
\label{fig:mcmc_straddle}
\end{figure*}

These differences are clearly shown in Figure \ref{fig:mcmc_straddle},
which depicts (with black dots) samples chosen by typical single runs of MCMC and
our algorithm when trying to compute the $95\%$ credible/confidence
intervals for a standard normal distribution\footnote{For the Bayesian case,
we assume that the observed data is a single point at the origin.  As
a result, the true posterior derived via sampling will be exactly
the same as the true standard Normal distribution.  This is done to
ensure that both algorithms are sampling the same function, allowing
us to compare the sampling patterns of the algorithms.}.
Both algorithms were
constrained to samples chosen in $[-10:10]$. The MCMC algorithm was
started at a randomly selected point, with a uniform prior over the
range. In this figure we use a standard normal proposal distribution,
although the sampling pattern is similar for other distributions we
tried.  Credible intervals for MCMC and confidence intervals
for our algorithm are depicted below the plots.  
Several points are quite apparent.  First
MCMC has failed to converge in 50 samples, while our algorithm has
converged nicely.  The credible intervals given by MCMC are not only
underestimated, but are also not centered on the true distribution's
center, revealing a potential liability for interpreting MCMC chains
which have not converged.

Secondly, notice that MCMC heavily samples the peak
of the distribution, while our algorithm focus on those regions
associated with the confidence interval boundaries.  The
MCMC chain results in a ragged collection of disjoint credible
intervals, while our algorithm returns a single interval in
which the endpoints have been well determined.

Thirdly, note that our algorithm samples extreme points to ensure that
it has not failed to observe additional peaks in the distribution
which may contribute to the 95\% confidence interval, while MCMC has
not. As noted before, since MCMC is not a search algorithm, it may
spend a large number of samples in a single distribution peak
before jumping to another peak in the distribution.  This sampling
pattern may cause MCMC to appear to have converged, when in reality
it has just failed to transition to the second peak, as in the two
Gaussian case described previously.

Finally, we note that the MCMC algorithm is not data efficient.  While
Figure \ref{fig:mcmc_straddle} depicts those experiments run by MCMC,
the final MCMC chain consists of only those points that were accepted
(in this case by the Metropolis-Hastings algorithm). As such, some of the
points that MCMC samples are discarded immediately, and never used to
guide the chain in future steps, or to determine the $1-\alpha$
credible intervals.  In addition, many MCMC practitioners 
remove all but every $j$th sample point (for some integer $j$) to
ensure that the points in the chain are truly independent. This
significantly reduces data efficiency.

\label{sec:bayesvsfreq}
\subsection{Advantages of Frequentist Inference}

Often, non-statisticians are confused by differences between Bayesian
and frequentist techniques, and the advantages and limitations that
each maintains.  Particularly appealing with the Bayesian approach is
the fact that one is computing a posterior distribution over the
parameter space.  Thus, not only does one obtain $1-\alpha$ credible
intervals, but one gets a sense of where within the interval, the
true value is expected to be.  Frequentist approaches do not allow for
one to compute the probability that the true value is equal to some
particular parameter value.  While choosing one technique over the
other is a matter of personal statistical philosophy, we believe that
frequentist approaches hold  important advantages over their Bayesian
counterparts. 

First, any Bayesian technique requires that one assume a family of likelihood
functions and a prior distribution over the parameter space in order
to compute the posterior.  The resulting posterior is only as valid as
both the likelihood and the prior.  In many cases, a prior
distribution is unknown. In these cases, an ``uninformative prior,''
equivalent to a uniform distribution on some bounded range, is often
assumed.  However, such a prior is not uninformative.  In particular,
a uniform prior indicates that the practitioner believes that the true
distribution of the parameter is uniform, not unknown.  Moreover ``uninformative''
priors are parametrization dependent.  If we reformulate our 7D CMB
problem by replacing $\Omega_M$ with $H_0$, a uniform prior over the
original problem will not translate into a uniform prior over the
formulation including $H_0$, as $\Omega_M$ is inversely related to
$H_0$.

Secondly, any change to the prior invalidates the current results.  In
particular, even when one is using a uniform prior, 
merely changing parameter 
ranges will result in a different posterior with possibly different
$1-\alpha$ credible intervals.  Thus analyses, like those we performed
in \S \ref{sec:intervals} would have required us to recompute the
entire chain (or set of chains), an extremely expensive proposition,
or somehow approximate the difference.
Additionally, for Bayesian techniques, the prior should be independent of
the data, and hence it should not be changed after observing the
data.  By recomputing the posterior using a new prior (based upon a
previous posterior), we open ourselves to errors incurred due to
multiple hypothesis testing.  Moreover, it is a small step from such
repeated Bayesian inferences to data-dependent priors, which are
incoherent not Bayesian.  Hence, data-dependent priors do not benefit
from theoretical guarantees derived for Bayesian analyses, which
assume priors are chosen before any data is observed.

It is interesting to note that Table 
\ref{paramtable} denotes the final ranges of parameters
searched.  We initially started with the same parameter ranges as
\citep{tegmark2001}, but increased our ranges slightly to better
capture a secondary peak in confidence space (shown in Figure
\ref{fig:results2d}). Because of our frequentist based technique, we
can easily change the ranges being searched without re-running any of
the CMBFast models, or recomputing any of our current inferences.
This contrasts sharply with Bayesian techniques.

Finally, recall from \S \ref{sec:introduction} that Bayesian approaches
answer a fundamentally different question than do frequentist
approaches.  Frequentist approaches are concerned with deriving
procedures which will return confidence intervals that trap the
true value of a parameter in at least $1-\alpha$ of the cases in which the
procedure is used.
Bayesian methods are more interested in determining the
probability that a particular value of a parameter is chosen for the
given data set and prior.
While we can compute ``credible'' intervals for Bayesian methods by
choosing the minimum range of a parameter such that the enclosed
probability is equal to $1-\alpha$, these intervals do not necessary
correspond to those derived from using a frequentist approach.  In
particular, there is no guarantee that credible intervals will
contain the true value of the parameter in at least $1-\alpha$
fraction of the instances where the technique is applied.
Specifically, when the likelihood function of the model goes awry,
such as in cases of high-dimension, missing data, and/or
non-parametric models, the inference made using Bayesian methods will
be incorrect.  

This problem is particularly acute for high dimensions,
where $1-\alpha$ credible intervals might trap the true value of the
parameter close to zero percent of the time.  That is, if Bayesian
techniques are applied to a series of data sets, the
fraction of the resulting $1-\alpha$ credible intervals that contain the true
values of the parameter will be less than $1-\alpha$ and may be
significantly less that $1-\alpha$.   While we find this fact
disturbing, a Bayesian might be willing to trade off the fact that the
credible intervals usually will not  contain the truth
for the ability to compute a posterior distribution of likelihood over
parameter space (assuming some prior) and hence determine the 
probability of any given parameter setting.
As, \cite{olivestatistics} notes: ``to construct procedures with
guaranteed long run performance, such as confidence intervals, use
frequentist methods.''

%
%
%
%
%
%

\section{Conclusions} \label{sec:conclusion}

In this paper, we present a new technique to map confidence surfaces, and
show results on first-year WMAP data.  This method, utilizing
a non-parametric fit and confidence balls, allows for computing
simultaneously valid confidence intervals.  
Our technique is similar in spirit to the Bayesian methods, but
differs significantly in that it is a frequentist analysis with
\textit{simultaneous valid} coverage.   
Thus, the derived confidence intervals are valid
regardless of the values of the remaining parameters.  This is not the
case when a maximization or marginalization technique is used. 
While the use of confidence balls requires a search over the entire
parameter space akin to the integration required for Bayesian
techniques, we present an algorithm to efficiently compute regions of
parameter space which have confidence values above a specified
$1-\alpha$ threshold.  We present results of our algorithm and note
that they are similar to those derived using alternative statistical
methods.  While the WMAP power spectrum data alone is insufficient to
constrain any of the cosmological parameters, the addition of
a reasonable assumption on the Hubble constant, provides useful
cosmological insights.

We point out that the purpose of this paper is to present
a new statistical and computational technique to provide
frequentist confidence intervals on the cosmological parameters
using the WMAP Year 1 data. We are not 
arguing that the allowed parameter ranges shown in Figures
\ref{fig:results1d}, \ref{fig:results2d}, \ref{fig:results1d:h0} and 
\ref{fig:results2d:h0}
 are more accurate than those presented by the WMAP
team. The reason for this is two-fold: (1) the comparison
of credible (Bayesian) vs. valid (frequentist) parameter ranges
is non-trivial and outside the scope of this work and (2) we
use only the WMAP Year 1 data, while others have utilized
non-WMAP data in various ways to provide additional
constraints on the parameters.

Analysis of Figures \ref{fig:results1d} and \ref{fig:results2d} shows
that the one sigma confidence regions are similar to those found in
the literature using a variety of techniques (e.g. \cite{tegmark2001,
  spergel2003, spergel2006}).  Figures
\ref{fig:results1d} and \ref{fig:results2d}  illustrate that
the WMAP data alone is not sufficient to strongly constrain the
matter/energy budget for the Universe.  In particular, the constraint
on the Hubble constant, $H_0$, is so weak as to allow values between
15 and 300 at the two sigma level.

If we instead constrain $H_0$ to a more ``typical'' range of
$[60:75]$, we get much tighter constraints on \textit{all} parameters,
as shown in Figures \ref{fig:results1d:h0} and \ref{fig:results2d:h0}.  
Because we are using a frequentist confidence procedure, adding the
restriction does not affect the validity of the inference.  Moreover,
no additional CMBFast models must be computed to test this constraint,
illustrating the power of our statistical procedure.
Note that both Figures \ref{fig:results1d:h0} and
\ref{fig:results2d:h0} agree much better with the current estimates of the
cosmological matter/energy budget and strongly suggest
that $\Omega_\mathrm{Total} = 1$.

Moreover, as we show in \S \ref{sec:convergence}, CMBFast creates
temperature power spectra which are discontinuous in parameter space.
This discontinuity violates the smoothness assumption of the
underlying target function used by both our Gaussian 
process technique, as well as by MCMC.   This makes convergence
statements difficult to make.  However, we believe that the 1.2
million models run show reasonable convergence.  We believe that with additional
assumptions on CMBFast --- such as the maximum size of a discontinuity
--- we will be able to prove that our method converges in a reasonable
time frame. 

Additionally, we show that comparing CMBFast models to the WMAP year 1
temperate power spectrum data results in a multi-modal solution in
confidence space. We have detected at least two distinct confidence
regions in parameter space.  However, by adding assumptions on $n_s$,
we can eliminate the secondary peak, leading us to believe that the
secondary peak may not be visible in the WMAP third year data. 

In summary, we believe the proposed approach of using a non-parametric
fit to the data and confidence balls, coupled with a search algorithm
to find models in parameter space which fit our regressed estimate,
provides a robust and informative 
method for computing confidence intervals for cosmological
parameters.  In addition to merely computing intervals, our approach
has the ability to test various constraints without computing new
models or making assumptions about which models should be fit and
what the ranges of the parameter space should be.  We are working on
techniques to prove convergence of the algorithm, as well as the
incorporation of additional data sets to further constrain the
mass/energy budget of the Universe.

\acknowledgments
The authors would like to thank the referee for his/her valuable
suggestions and corrections.

{\it Facilities:} \facility{WMAP}
\appendix
\section{Estimating $\tau$} \label{appendix}

Recall from \S \ref{sec:fit} that the cosine basis is defined on
$[0,1]$ by
\[
\phi_j(x) =
\left\{
\begin{array}{l l}
1 & \mathrm{for\ } j=0\\
\sqrt{2}\cos(\pi j x) & \mathrm{for\ } j = 1,2,3, \dots
\end{array}
\right.
\]
If $j$ and $k$ are distinct, positive integers, then
\begin{eqnarray*}
\phi_j \phi_k &=& 2 \cos(\pi j x) \cos(\pi k x)\\ &=& \cos(\pi(j+k)x) + \cos(\pi(j-k)x)\\
&=& \frac{1}{\sqrt{2}} (\phi_{j+k} + \phi_{|j-k|}).
\end{eqnarray*}
Moreover, if $j>0$, then $\phi_j^2 = 2 \cos^2(\pi j x) = \cos(2\pi j x)+1  = \frac{1}{\sqrt{2}}
\phi_{2j} + \phi_0.$
Therefore, as mentioned in \S \ref{sec:fit},
\[
\Delta_{jk\ell} = \left\{ 
\begin{array}{c c}
1 & \mathrm{if\ \#}\{j,k,l = 0\} = 3\\
0 & \mathrm{if\ \#}\{j,k,l = 0\} = 2\\
\delta_{jk}\delta_{0\ell} + \delta_{j\ell}\delta_{0k} +
\delta_{k\ell}\delta_{0j} & \mathrm{if\ \#}\{j,k,l = 0\} = 1\\
\frac{1}{\sqrt{2}}(\delta_{\ell, j+k} + \delta_{\ell,|j-k|}) & \mathrm{if\ \#}\{j,k,l = 0\} = 0
\end{array}
\right..
\]
Let $w(x) = 1/\sigma^2(x)$, such that $w^2(x) = \sum_j w_j
\phi_j(x)$.  As in \S \ref{sec:fit}, we let $\hat \mu_j = \lambda_j
Z_j$, where
\[
Z_j = \frac{1}{n} \sum_{i=1}^n Y_i \phi_j(X_i)
\]
and $1 \ge \lambda_0 \ge \lambda_1 \ge \cdots \ge \lambda_n \ge 0$ are
shrinkage coefficients.  In this work, we use a special case of
monotone shrinkage in which 
\[
\lambda_j = \left\{ 
\begin{array}{cc}
1 & \mathrm{for\ } j\le J\\
0 & \mathrm{for\ } j> J
\end{array}\right.
\]
for $J \in [0,1,2,\dots,n]$ such that $J$ minimizes Stein's unbiased
risk estimate given in Equation \ref{eqn:stein}.
With these definitions, the loss can be written as
\begin{eqnarray*}
L(f, \hat f) 
&=& 
\int_0^1 \left(\frac{\hat f(x) -
  f(x)}{\sigma(x)}\right)^2 \, dx\\
&=&
\sum_{j,k,\ell} (\mu_j - \hat \mu_j)(\mu_k - \hat \mu_k) w_\ell
\int_0^1 \phi_j \phi_k \phi_\ell\\
&=&
\sum_{j,k} (\mu_j - \hat \mu_j)(\mu_k - \hat \mu_k) 
\sum_{\ell} w_\ell \Delta_{jk\ell}\\
&=& (\mu - \hat \mu)^T W (\mu - \hat \mu),
\end{eqnarray*}
where $W_{jk} = \sum_\ell w_\ell \Delta_{jk\ell}$.  
As in \S \ref{sec:fit}, let $D$ and $\bar D = 1 -D$ be diagonal matrices with 1's in the
first $J$ and last $n-J$ entries respectively.  Then $\hat \mu = DZ$,
where $Z$ is again assumed to be Normal $(\mu, B)$.  Thus,
$\E [\hmu] = D \mu$, $\Cov(\hmu_j, \hmu_k) = \lambda_j\lambda_k B_{jk}$
and $\Var(\hmu) = DBD$.  The risk then becomes
\ba
R = \E [L] &=& \E \left[(\mu - \hmu)^T W (\mu - \hmu)\right]\\
&=& \mathrm{trace}(DWDB)+ \mu^T\bar D W \bar D \mu\\
&=& \mathrm{trace}(DWDB)+ \sum_{j,k} \mu_j \mu_k \bar \lambda_j \bar
\lambda_k W_{jk}
\ea
An unbiased estimate can be obtained by replacing $\mu_j \mu_k$ with
$Z_j Z_k - B_{jk}$.  The result is
\[
\hat R = Z^T \bar D W \bar D Z + \mathrm{trace}(DWDB) -
\mathrm{trace}(\bar D W \bar D B)
\]
It follows that
\[
\hat L - \hat R =
\mu^T W \mu -Z^TC + Z^T A Z + \mathrm{trace}(AZ)
\]
where $A = DW+WD-W$ and $C = 2DW\mu$.  Moreover,
\ba
\Var(\hat L - \hat R)  &=& \Var(Z^T A Z - Z^TC)\\
&=& \Var(Z^T A Z) + \Var(Z^T C) - 2\,\Cov(Z^TAZ, Z^TC)\\
&=&2 \, \mathrm{trace}(ABAB) + \mu^T Q \mu
\ea
where $Q = ABA + WDBDW - 2ABDW$.  Plugging in unbiased estimates of
the linear and quadratic forms involving $\mu$, we get the following
estimate for the variance of the pivot process:
\[
\hat \tau^2 = 2\, \mathrm{trace}(ABAB)+ Z^TQZ - \mathrm{trace}(QB).
\]

\bibliographystyle{apj}
\bibliography{ms}

\begin{thebibliography}{38}
\expandafter\ifx\csname natexlab\endcsname\relax\def\natexlab#1{#1}\fi

\bibitem[{{Abroe} {et~al.}(2002){Abroe}, {Balbi}, {Borrill}, {Bunn}, {Hanany},
  {Ferreira}, {Jaffe}, {Lee}, {Olive}, {Rabii}, {Richards}, {Smoot}, {Stompor},
  {Winant}, \& {Wu}}]{abroe2002}
{Abroe}, M.~E., {Balbi}, A., {Borrill}, J., {Bunn}, E.~F., {Hanany}, S.,
  {Ferreira}, P.~G., {Jaffe}, A.~H., {Lee}, A.~T., {Olive}, K.~A., {Rabii}, B.,
  {Richards}, P.~L., {Smoot}, G.~F., {Stompor}, R., {Winant}, C.~D., \& {Wu},
  J.~H.~P. 2002, \mnras, 334, 11

\bibitem[{{Bennett} {et~al.}(2003){Bennett}, {Hill}, {Hinshaw}, {Nolta},
  {Odegard}, {Page}, {Spergel}, {Weiland}, {Wright}, {Halpern}, {Jarosik},
  {Kogut}, {Limon}, {Meyer}, {Tucker}, \& {Wollack}}]{bennett2003}
{Bennett}, C.~L., {Hill}, R.~S., {Hinshaw}, G., {Nolta}, M.~R., {Odegard}, N.,
  {Page}, L., {Spergel}, D.~N., {Weiland}, J.~L., {Wright}, E.~L., {Halpern},
  M., {Jarosik}, N., {Kogut}, A., {Limon}, M., {Meyer}, S.~S., {Tucker}, G.~S.,
  \& {Wollack}, E. 2003, \apjs, 148, 97

\bibitem[{Beran(2000)}]{beran2000}
Beran, R. 2000, Journal of the American Statistical Association, 95, 155

\bibitem[{Beran \& D\"umbgen(1998)}]{beran1998}
Beran, R. \& D\"umbgen, L. 1998, Annuals of Statistics, 26, 1826

\bibitem[{Bryan {et~al.}(2005)Bryan, Schneider, Nichol, Miller, Genovese, \&
  Wasserman}]{bryan2005}
Bryan, B., Schneider, J., Nichol, R.~C., Miller, C.~J., Genovese, C.~R., \&
  Wasserman, L. 2005, in Advances in Neural Information Processing Systems 18
  (Cambridge, MA: MIT Press)

\bibitem[{Christensen {et~al.}(2001)Christensen, Meyer, Knox, \&
  Luey}]{christensen2001}
Christensen, N., Meyer, R., Knox, L., \& Luey, B. 2001, Classical and Quantum
  Gravity, 18, 2677

\bibitem[{{Chu} \& {Knox}(2005)}]{chu2005}
{Chu}, M. \& {Knox}, L. 2005, \apj, 620, 1

\bibitem[{Cressie(1991)}]{cressie1991}
Cressie, N. A.~C. 1991, Statistics for Spatial Data (New York: Wiley)

\bibitem[{{Dunkley} {et~al.}(2005){Dunkley}, {Bucher}, {Ferreira}, {Moodley},
  \& {Skordis}}]{dunkley2005}
{Dunkley}, J., {Bucher}, M., {Ferreira}, P.~G., {Moodley}, K., \& {Skordis}, C.
  2005, \mnras, 356, 925

\bibitem[{{Fendt} \& {Wandelt}(2006)}]{fendt2006}
{Fendt}, W.~A. \& {Wandelt}, B.~D. 2006, ArXiv Astrophysics e-prints

\bibitem[{{Genovese} {et~al.}(2004){Genovese}, {Miller}, {Nichol},
  {Arjunwadkar}, \& {Wasserman}}]{genovese2004}
{Genovese}, C., {Miller}, C.~J., {Nichol}, R.~C., {Arjunwadkar}, M., \&
  {Wasserman}, L. 2004, Statistic Science, 19, 308

\bibitem[{{Gorski} {et~al.}(1993){Gorski}, {Stompor}, \&
  {Juszkiewicz}}]{gorski1993}
{Gorski}, K.~M., {Stompor}, R., \& {Juszkiewicz}, R. 1993, \apjl, 410, L1

\bibitem[{{Griffiths} {et~al.}(2001){Griffiths}, {Silk}, \&
  {Zaroubi}}]{griffiths2001}
{Griffiths}, L.~M., {Silk}, J., \& {Zaroubi}, S. 2001, \mnras, 324, 712

\bibitem[{Guestrin {et~al.}(2005)Guestrin, Krause, \& Singh}]{guestrin2005}
Guestrin, C., Krause, A., \& Singh, A.~P. 2005, in ICML '05: Proceedings of the
  22nd international conference on Machine learning (New York, NY, USA: ACM
  Press), 265--272

\bibitem[{{Gupta} \& {Heavens}(2002)}]{gupta2002}
{Gupta}, S. \& {Heavens}, A.~F. 2002, \mnras, 334, 167

\bibitem[{{Hajian}(2006)}]{hajain2006}
{Hajian}, A. 2006, ArXiv Astrophysics e-prints

\bibitem[{Hart {et~al.}(1968)Hart, Nilsson, \& Raphael}]{hart1968}
Hart, P., Nilsson, N., \& Raphael, B. 1968, in IEEE Trans. on Systems Science
  and Cybernetics, Vol.~4, IEEE, 100--107

\bibitem[{{Hinshaw} {et~al.}(2003){Hinshaw}, {Spergel}, {Verde}, {Hill},
  {Meyer}, {Barnes}, {Bennett}, {Halpern}, {Jarosik}, {Kogut}, {Komatsu},
  {Limon}, {Page}, {Tucker}, {Weiland}, {Wollack}, \& {Wright}}]{hinshaw2003}
{Hinshaw}, G., {Spergel}, D.~N., {Verde}, L., {Hill}, R.~S., {Meyer}, S.~S.,
  {Barnes}, C., {Bennett}, C.~L., {Halpern}, M., {Jarosik}, N., {Kogut}, A.,
  {Komatsu}, E., {Limon}, M., {Page}, L., {Tucker}, G.~S., {Weiland}, J.~L.,
  {Wollack}, E., \& {Wright}, E.~L. 2003, \apjs, 148, 135

\bibitem[{{Jimenez} {et~al.}(2004){Jimenez}, {Verde}, {Peiris}, \&
  {Kosowsky}}]{jimenez2004}
{Jimenez}, R., {Verde}, L., {Peiris}, H., \& {Kosowsky}, A. 2004, \prd, 70,
  023005

\bibitem[{{Kaplinghat} {et~al.}(2002){Kaplinghat}, {Knox}, \&
  {Skordis}}]{kaplinghat2002}
{Kaplinghat}, M., {Knox}, L., \& {Skordis}, C. 2002, \apj, 578, 665

\bibitem[{{Knox} {et~al.}(2001){Knox}, {Christensen}, \& {Skordis}}]{knox2001}
{Knox}, L., {Christensen}, N., \& {Skordis}, C. 2001, \apjl, 563, L95

\bibitem[{{Lewis} \& {Bridle}(2002)}]{lewis2002}
{Lewis}, A. \& {Bridle}, S. 2002, \prd, 66, 103511

\bibitem[{{Lewis} {et~al.}(2000){Lewis}, {Challinor}, \& {Lasenby}}]{lewis2000}
{Lewis}, A., {Challinor}, A., \& {Lasenby}, A. 2000, \apj, 538, 473

\bibitem[{MacKay(1992)}]{mackay1992}
MacKay, D. 1992, Neural Computation, 4, 590

\bibitem[{{Miller} {et~al.}(2001){Miller}, {Nichol}, {Genovese}, \&
  {Wasserman}}]{miller2001}
{Miller}, C.~J., {Nichol}, R.~C., {Genovese}, C., \& {Wasserman}, L. 2001,
  Bulletin of the American Astronomical Society, 33, 1358

\bibitem[{{Padmanabhan} \& {Sethi}(2001)}]{padmanabhan2001}
{Padmanabhan}, T. \& {Sethi}, S.~K. 2001, \apj, 555, 125

\bibitem[{{Sandvik} {et~al.}(2004){Sandvik}, {Tegmark}, {Wang}, \&
  {Zaldarriaga}}]{sandvik2004}
{Sandvik}, H.~B., {Tegmark}, M., {Wang}, X., \& {Zaldarriaga}, M. 2004, \prd,
  69, 063005

\bibitem[{Schafer \& Stark(2003)}]{schafer2003}
Schafer, C.~M. \& Stark, P.~B. 2003, in Proceedings for Statistical Problems in
  Particle Physics, Astrophysics, and Cosmology, SLAC

\bibitem[{{Seljak} \& {Zaldarriaga}(1996)}]{seljak1996}
{Seljak}, U. \& {Zaldarriaga}, M. 1996, Astrophyical Journal, 469, 437

\bibitem[{Seljak {et~al.}(2003)Seljak, Sugiyama, White, \&
  Zaldarriaga}]{seljak2003}
Seljak, U. c.~v., Sugiyama, N., White, M., \& Zaldarriaga, M. 2003, Phys. Rev.
  D, 68, 083507

\bibitem[{Spergel {et~al.}(2006)Spergel, Bean, Dore, Nolta, Bennett, Hinshaw,
  Jarosik, Komatsu, Page, Peirisand, Verde, Barnes, Halpern, Hill, Kogut,
  Limon, Meyer, Odegard, Tucker, Weiland, Wollack, \& Wright}]{spergel2006}
Spergel, D.~N., Bean, R., Dore, O., Nolta, M.~R., Bennett, C.~L., Hinshaw, G.,
  Jarosik, N., Komatsu, E., Page, L., Peirisand, H.~V., Verde, L., Barnes, C.,
  Halpern, M., Hill, R.~S., Kogut, A., Limon, M., Meyer, S.~S., Odegard, N.,
  Tucker, G.~S., Weiland, J.~L., Wollack, E., \& Wright, E.~L. 2006,
  astro-ph/0603449

\bibitem[{{Spergel} {et~al.}(2003){Spergel}, {Verde}, {Peiris}, {Komatsu},
  {Nolta}, {Bennett}, {Halpern}, {Hinshaw}, {Jarosik}, {Kogut}, {Limon},
  {Meyer}, {Page}, {Tucker}, {Weiland}, {Wollack}, \& {Wright}}]{spergel2003}
{Spergel}, D.~N., {Verde}, L., {Peiris}, H.~V., {Komatsu}, E., {Nolta}, M.~R.,
  {Bennett}, C.~L., {Halpern}, M., {Hinshaw}, G., {Jarosik}, N., {Kogut}, A.,
  {Limon}, M., {Meyer}, S.~S., {Page}, L., {Tucker}, G.~S., {Weiland}, J.~L.,
  {Wollack}, E., \& {Wright}, E.~L. 2003, \apjs, 148, 175

\bibitem[{{Tegmark}(1999)}]{tegmark1999}
{Tegmark}, M. 1999, \apjl, 514, L69

\bibitem[{{Tegmark} {et~al.}(2004){Tegmark}, {Strauss}, {Blanton}, {Abazajian},
  {Dodelson}, {Sandvik}, {Wang}, {Weinberg}, {Zehavi}, {Bahcall}, {Hoyle},
  {Schlegel}, {Scoccimarro}, {Vogeley}, {Berlind}, {Budavari}, {Connolly},
  {Eisenstein}, {Finkbeiner}, {Frieman}, {Gunn}, {Hui}, {Jain}, {Johnston},
  {Kent}, {Lin}, {Nakajima}, {Nichol}, {Ostriker}, {Pope}, {Scranton},
  {Seljak}, {Sheth}, {Stebbins}, {Szalay}, {Szapudi}, {Xu}, {Annis},
  {Brinkmann}, {Burles}, {Castander}, {Csabai}, {Loveday}, {Doi}, {Fukugita},
  {Gillespie}, {Hennessy}, {Hogg}, {Ivezi{\'c}}, {Knapp}, {Lamb}, {Lee},
  {Lupton}, {McKay}, {Kunszt}, {Munn}, {O'Connell}, {Peoples}, {Pier},
  {Richmond}, {Rockosi}, {Schneider}, {Stoughton}, {Tucker}, {vanden Berk},
  {Yanny}, \& {York}}]{tegmark2004}
{Tegmark}, M., {Strauss}, M.~A., {Blanton}, M.~R., {Abazajian}, K., {Dodelson},
  S., {Sandvik}, H., {Wang}, X., {Weinberg}, D.~H., {Zehavi}, I., {Bahcall},
  N.~A., {Hoyle}, F., {Schlegel}, D., {Scoccimarro}, R., {Vogeley}, M.~S.,
  {Berlind}, A., {Budavari}, T., {Connolly}, A., {Eisenstein}, D.~J.,
  {Finkbeiner}, D., {Frieman}, J.~A., {Gunn}, J.~E., {Hui}, L., {Jain}, B.,
  {Johnston}, D., {Kent}, S., {Lin}, H., {Nakajima}, R., {Nichol}, R.~C.,
  {Ostriker}, J.~P., {Pope}, A., {Scranton}, R., {Seljak}, U., {Sheth}, R.~K.,
  {Stebbins}, A., {Szalay}, A.~S., {Szapudi}, I., {Xu}, Y., {Annis}, J.,
  {Brinkmann}, J., {Burles}, S., {Castander}, F.~J., {Csabai}, I., {Loveday},
  J., {Doi}, M., {Fukugita}, M., {Gillespie}, B., {Hennessy}, G., {Hogg},
  D.~W., {Ivezi{\'c}}, {\v Z}., {Knapp}, G.~R., {Lamb}, D.~Q., {Lee}, B.~C.,
  {Lupton}, R.~H., {McKay}, T.~A., {Kunszt}, P., {Munn}, J.~A., {O'Connell},
  L., {Peoples}, J., {Pier}, J.~R., {Richmond}, M., {Rockosi}, C., {Schneider},
  D.~P., {Stoughton}, C., {Tucker}, D.~L., {vanden Berk}, D.~E., {Yanny}, B.,
  \& {York}, D.~G. 2004, \prd, 69, 103501

\bibitem[{{Tegmark} {et~al.}(2001){Tegmark}, {Zaldarriaga}, \&
  {Hamilton}}]{tegmark2001}
{Tegmark}, M., {Zaldarriaga}, M., \& {Hamilton}, A.~J. 2001, Physical Review D,
  63

\bibitem[{{Verde} {et~al.}(2003){Verde}, {Peiris}, {Spergel}, {Nolta},
  {Bennett}, {Halpern}, {Hinshaw}, {Jarosik}, {Kogut}, {Limon}, {Meyer},
  {Page}, {Tucker}, {Wollack}, \& {Wright}}]{verde2003}
{Verde}, L., {Peiris}, H.~V., {Spergel}, D.~N., {Nolta}, M.~R., {Bennett},
  C.~L., {Halpern}, M., {Hinshaw}, G., {Jarosik}, N., {Kogut}, A., {Limon}, M.,
  {Meyer}, S.~S., {Page}, L., {Tucker}, G.~S., {Wollack}, E., \& {Wright},
  E.~L. 2003, \apjs, 148, 195

\bibitem[{Wasserman(2004)}]{olivestatistics}
Wasserman, L. 2004, All of Statistics (New York: Springer-Verlag)

\bibitem[{{White} \& {Bunn}(1995)}]{white1995}
{White}, M. \& {Bunn}, E.~F. 1995, \apj, 450, 477

\end{thebibliography}
\end{document}